\documentclass[aps,prb,showpacs,twocolumn]{revtex4-1}
\usepackage{bm,color,amsmath,amssymb,mathrsfs,latexsym,graphicx,psfrag}

\def\ea{{\it et al.}}

\newcommand{\bk}{{\mathbf k}}

\newcommand{\br}{{\mathbf r}}

\newcommand{\be}{\begin{equation}}
\newcommand{\ee}{\end{equation}}

\def\be{\begin{equation}}
\def\ee{\end{equation}}
\def\bea{\begin{eqnarray}}
\def\eea{\end{eqnarray}}

\def\C60{A$_x$C$_{60}$}

\def\HgCu3{HgCa$_2$Cu$_3$O$_{8+y}$}
\def\HgCu4{HgBa$_2$Ca$_3$Cu$_4$O$_{10+y}$}
\def\TlCu{Tl$_2$Ba$_2$CuO$_{6+\delta}$}
\def\TlCu3{Tl$_2$Ba$_2$Ca$_2$Cu$_3$O$_{10+y}$}
\def\TlCu4{Tl$_2$Ba$_2$Ca$_3$Cu$_4$O$_{12+y}$}

\def\BiCu3{Bi$_2$Sr$_2$Ca$_{2}$Cu$_3$O$_y$}

\def\8LSCO{La$_{1.88}$Sr$_{.12}$CuO$_4$}
\def\110LNSCO{La$_{1.5}$Nd$_{0.4}$Sr$_{0.1}$CuO$_{4}$}
\def\stage4LCO{La$_{2}$CuO$_{4+\delta}$}
\def\Y248{YBa$_2$Cu$_4$O$_8$}

\def\NbSe2{NbSe$_2$}
\def\TaSe2{TaSe$_2$}
\def\TiSe2{TiSe$_2$}

\begin{document}

\title{Entanglement Spectrum Classification of $C_n$-invariant Noninteracting Topological Insulators in Two Dimensions}
\author{Chen Fang$^{1}$, Matthew J. Gilbert$^{2,3}$,  B. Andrei Bernevig$^{1}$}
\affiliation{$^1$Department of Physics, Princeton University, Princeton NJ 08544}
\affiliation{$^2$Department of Electrical and Computer Engineering, University of Illinois,  Urbana IL 61801, USA}
\affiliation{$^3$Micro and Nanotechnology Laboratory, University of Illinois, 208 N. Wright St, Urbana IL 61801, USA}
\date{\today}

\begin{abstract}
We study the single particle entanglement spectrum in 2D topological insulators which possess $n$-fold rotation symmetry. By defining a series of special choices of subsystems on which the entanglement is calculated, or real space cuts, we find that the number of protected in-gap states for each type of these real space cuts is a quantum number indexing (if any) non-trivial topology in these insulators. We explicitly show the number of protected in-gap states is determined by a $Z^n$-index, $(z_1,...,z_n)$, where $z_m$ is the number of occupied states that transform according to $m$-th one-dimensional representation of the $C_n$ point group. We find that the entanglement spectrum contains in-gap states pinned in an interval of entanglement eigenvalues $[1/n,1-1/n]$. We determine the number of such in-gap states for an exhaustive variety of cuts, in terms of the $Z_m$ quantum numbers. Furthermore, we show that in a homogeneous system, the $Z^n$ index can be determined through an evaluation of the eigenvalues of point group symmetry operators at all high-symmetry points in the Brillouin zone. When disordered $n$-fold rotationally symmetric systems are considered, we find that the number of protected in-gap states is identical to that in the clean limit as long as the disorder preserves the underlying point group symmetry and does not close the bulk insulating gap.
\end{abstract}
\maketitle

The study of novel topological phases of matter has become one of the most active fields in condensed matter physics\cite{qi2005,Kane:2005sf,Bernevig:2006kx,Fu:2006rm,moore2007,koenig2007,Fu:2007fk,Hsieh:2008fk,hsieh2009a,zhang2009,liu2009,ZhouX2009,YaoH2010,chiralTSQH,qi2011rev,SunK2011,Wan2011,XuG2011,Fang:2011,HalaszG2012}. Chronologically speaking, the first of these phases to be experimentally realized is the integer quantum Hall (IQH) state, which is a striking departure from the traditional theory of conductivity due to its quantized Hall conductance and chiral edge modes\cite{Klitzing}. Shortly after its experimental discovery, Thouless demonstrated that the special properties of the IQH state came down to its non-trivial bandstructure topology\cite{Thouless:1982rz}.
For the IQH state, one can prove that the quantized Hall conductance is equivalent to a momentum space integral of the Berry curvature, which is a non-zero integer (known as the Chern number in topology), multiplied by the conductance quantum $e^2/h$. Beyond this, Haldane demonstrated that the IQH state may be further generalized to a system which does not require an external overall magnetic flux, yet still possesses a non-zero Chern number and quantized Hall conductance. This system is generally referred to as a Chern insulator\cite{Haldane1988}.

Further as systems which preserve time-reversal invariance (TRI) pose an interesting problem as their Chern number vanishes in the presence of non-trivial topology. This necessitates the definition of a new quantum number capable of distinguishing between trivial and non-trivial topological states in systems which possess TRI. In 2D\cite{qi2005,Bernevig:2006kx,Kane:2005sf} and 3D\cite{Fu:2007fk,Qi:2008sf}, one can define a $Z_2$-number,  which here we call $\gamma_0$, as a quantity which is uniquely determined by the matrix representations of the time-reversal symmetry operator at all time-reversal invariant points within the Brillouin zone (BZ). In gapped systems, $\gamma_0=1$ corresponds to topologically non-trivial insulators while $\gamma_0=0$ corresponds to trivial insulators. In real materials, TRI topologically non-trivial insulators, or simply topological insulators usually have strong spin-orbit interactions which lead to an inversion of the bulk band gap at an odd number of time-reversal invariant points within the BZ\cite{hasan2010,qi2011rev}.

The important difference between Chern insulators (characterized by $Z$-indices) and TRI topological insulators (characterized by $Z_2$-indices) is the absence/presence of time-reversal symmetry. In general, we expect that the dimensionality and the symmetry of the system will determine the available quantum numbers that may be used to distinguish the topologically non-trivial states from the trivial ones. Motivated by this notion, Schnyder \ea~classified translationally invariant topological insulators in presence/absence of three global symmetries: time-reversal, particle-hole, and chiral in 2D and 3D\cite{schnyder2008,Ryu2010}. With translational symmetry, a non-interacting insulator maps to a manifold defined on the entire BZ. At each $\bk$ in BZ, the Hamiltonian maps to a projector onto the subspace spanned by the occupied Bloch states, or an element of the group $U(N_{orb})/U(N_{occ})\times{U}(N_{orb}-N_{occ})$. The mathematical mappings of various insulators have been previously studied and it is found that they may be classified by their specific homotopy groups\cite{nakahara1990}. To be more concrete, in 2D the homotopy group for systems with only translational symmetry is $\pi_2(U(N_{orb})/U(N_{occ})\times{U}(N_{orb}-N_{occ}))=Z$, which indicates that topological phases are indexed by integers. This picture is consistent with the previously understood conclusion that 2D systems may be characterized by a Chern number. In the same manner, TRI 2D and 3D insulators are also classified by the corresponding homotopy group, $Z_2$, of continuous mappings from the $d=2,3$-dimensional BZ to a target space of projectors to occupied states under the constraint of TRI. Besides the bulk topological indices defined in insulators with translational invariance and with disorder, a bulk-edge correspondence exists which guarantees presence of gapless modes on the  boundary between two insulators having different values of $Z$ or $Z_2$ index\cite{hasan2010,qi2006,chandran2011}.

The three symmetries considered above are local symmetries that do not involve spatial degrees of freedom. In condensed matter systems, we are faced with many types of symmetry operations in real space, besides time-reversal, charge conjugation and chiral symmetries. Amongst these symmetries, crystallographic point group symmetries (PGS) are well-known to universally exist within solids and may thus provide a natural avenue through which to continue the expansion of the family of materials which harbor topologically non-trivial states. Along this line, Fu\cite{Fu:2011} first studied spinless 3D insulators exhibiting gapless surface modes with quadratic band dispersion, the existence of which is protected by $C_{4,6}$ rotation PGS and TRI. Furthermore, within the context of cold atoms, Sun \ea~discussed a quadratic band crossing point in 2D that is protected by $C_4$ PGS and TRI. These examples provide a tantalizing glimpse into the possibilities which exist for finding new non-trivial states of matter when PGS is combined with TRI. An equally interesting question one may ask is:  can PGS bring about new topological phases, \emph{without} the presence of TRI and how shall we identify them\cite{Wan2011,Fang:2011}? Without the presence of TRI, we are no longer \emph{a priori} guaranteed to have any of the topological characteristics which are hallmarks of this symmetry such as: quantized magneto-electric coefficient or boundary/surface modes thereby necessitating the definition of a new topological quantum number.

To this end, the study of insulators with inversion symmetry\cite{hughes2010inv,Alexandradinata:2012,Turner:2012}, which is the simplest yet non-trivial PGS, shows that in these insulators a new topological number may indeed be defined which is independent of the $Z$- or $Z_2$ numbers previously introduced. This number can be used to characterize an inversion invariant insulator which contains a null result for both the Chern number ($Z$-number) and $Z_2$-number, but is still topologically distinct from a trivial insulator. The new quantum number labeling the new topological insulators is the number of protected in-gap states in the single particle entanglement spectrum.

The entanglement spectrum has been successfully applied to identify topological orders in a variety of condensed matter systems such as fractional quantum Hall systems\cite{li2008}, spin chains\cite{thomale2010}, Chern insulators\cite{prodan2010}, and topological band insulators\cite{turner2010,chandran2011,pollmann2010,fidkowski2010}. Unlike the energy spectrum, the quantum entanglement spectrum solely depends on the many-body ground state, and is, for this reason, more suitable for identifying any non-trivial topology in the ground state. In noninteracting systems, it can be shown that to obtain the many-body entanglement spectrum, only the single particle entanglement spectrum is needed, defined as all eigenvalues of the one-body reduced density matrix\cite{Peschel1993,Alexandradinata2011}:\bea\label{eq:DefEntSpec} C_{ij}(A)=\langle{c}^\dag_ic_j\rangle,\eea where $i,j\in{A}$ and $A$ is a sub-system of the whole system $L$. In principle, all `energies' in the entanglement spectrum are between zero and unity corresponding to states which are either localized outside of or inside the subsystem, respectively. In the case of inversion symmetric topological insulators\cite{hughes2010inv}, if $A$ is chosen as the left/right half of the $L$, then there can be several eigenstates of $C(A)$ with degenerate eigenvalue at exact $1/2$, in an inversion invariant insulator. These degenerate eigenvalues correspond to states that are topologically protected in the sense that they always stay at pinned $1/2$ so long as the inversion symmetry is preserved and the bulk gap remains open. These states are referred to as the protected $1/2$-in-gap states and their number defines a new topological number to be used to differentiate between trivial and non-trivial inversion invariant insulators.

In this work, we focus on insulators with another type of PGS, $n$-fold rotation symmetry where $n=2,3,4,6$ by lattice restriction. We show that for proper choices of sub-system $A$ in PGS invariant insulators, there exist protected in-gap states the number of which defines a new topological invariant. While these states are analogous to the $1/2$-in-gap states previously seen in inversion invariant insulators, two key differences are noted: first, instead of taking $A$ as the left/right half of $L$, there exists more than one type of symmetric cuts for a $C_n$ invariant insulator and, in general, each has its own corresponding number of in-gap states. Second, the protected in-gap states now are not necessarily located at exact $1/2$, but within a region around $1/2$, prevented by symmetry from arbitrarily approaching two ends of the single particle entanglement spectrum (zero and unity). We calculate the number of protected in-gap states for each type of symmetric cuts in a $C_n$ invariant insulator in terms of $m$-fold rotation eigenvalues at high-symmetry points in the BZ, where $m$ is a factor of $n$. Further, we show that all the numbers of protected in-gap states can be expressed in terms of a $Z^n$-index $(z_1,z_2,...,z_n)$, giving a $Z^n$-classification of these insulators. The physical reason underlying the $Z^n$ classification is very simple: if the single particle Hamiltonian is $n$-fold invariant, then every single particle wavefunction belongs to a certain 1D representation of $C_n$ group. The number of electrons in each representation below the Fermi level then naturally gives the index $(z_1,z_2,...,z_n)$. Finally, we will show that if the translational symmetry is broken by introducing disorder potential that is also $C_n$ invariant, the $Z^n$-index remains unchanged and as do the associated numbers of protected in-gap states.

The paper is organized as follows. In Sec.\ref{sec:prelim}, we introduce concepts related to our examination of PGS insulators in 2D referred to throughout the paper:  In Sec.\ref{sec:prelim}(A), we introduce crystallographic point groups; in in Sec.\ref{sec:prelim}{B}, we discuss a general definition of non-interacting topologically non-trivial insulator, and in Sec.\ref{sec:prelim}(C), a brief introduction to single particle entanglement spectrum and its relation to entanglement entropy. In Sec.\ref{sec:1Dinv} we revisit the entanglement spectrum in a 1D inversion invariant system. In this revisit, we analytically prove the relation between the number of protected in-gap states and the inversion eigenvalues at $k=0$ and $k=\pi$. The emphasis here is the relation between the number of $1/2$-in-gap states and a $Z^2$-index $(z_1,z_2)$, where $z_1$ ($z_2$) is the number of occupied Bloch states with odd (even) parity. Sec.\ref{sec:2D3Dinv} we perform a simple extension of the previous 1D results to 2D and 3D inversion invariant insulators, where we relate the number of $1/2$-in-gap states to the parity of Chern number and, when considering systems with TRI, to the $Z_2$ number. In Sec.\ref{sec:Cn}, we present our main result. We outline the analytic proof by which we relate the $Z^n$ index to the number of protected in-gap states in a $C_n$ invariant insulator with every type of possible symmetric cut, with some specific details relegated to the various Appendices. In Sec.\ref{sec:discuss}, we briefly discuss 3D PGS invariant insulators in Sec.\ref{sec:discuss}(A); we discuss in Sec.\ref{sec:discuss}(B) the effect of weak interaction, which reduces the $Z^n$-index down to a $Z_n$-index as single particle states lose their meaning while the \emph{many-body} wavefunction of any non-degenerate insulating state is still a 1D representation and can be associated with some $Z_{n}$ number.

\section{Preliminaries}
\label{sec:prelim}

Before proceeding with the relevant background material, we briefly clarify the notations we use throughout the paper. When we use the notation $C_n$, we are referring to either the point group $C_n$, or to the $n$-fold rotation, the choice of which should be clear given the context. $\hat{C}_n$ refers an operator in the Hilbert space corresponding to the $n$-fold rotation. $\mathcal{C}_n$ denotes the matrix representation of the operator $\hat{C}_n$ in a basis, either the orbital basis or the occupied-band basis, which should be discernible given the context of its usage. Similar definitions and distinctions apply to other operators, e.g., $M_z$, $\hat{M}_z$, $\mathcal{M}_z$ refer to mirror symmetry $z\rightarrow{-z}$, its operator in Hilbert space, and a matrix representation of the operator, respectively. Finally, when used as indices, the Greek letters will always span the orbitals, while Roman letters denote either the bands or spatial directions ($x,y,z$ or $k_x,k_y,k_z$). In all equations, repeated indices are, unless otherwise noted, automatically summed for the following types of summations: summation over all sites in the system, summation over all sites in the subsystem on which the entanglement spectrum is calculated, summation over all orbitals, summation over all bands (occupied and unoccupied).

\subsection{2D Crystallographic Point Groups}
A point group is a group of all symmetry operations (an atom, a molecule or a lattice) that leave at least one point fixed in space. In nature, there exist an infinite number of point groups. On the other hand, crystallographic point groups are point groups that are consistent with lattice translational symmetry. The crystallographic restriction theorem states that lattice translational symmetry dictates that $n=1,2,3,4,6$ for any $n$-fold rotation axis and effectively limits the total number of crystallographic point groups. In 2D there are 10 different point groups allowed which are comprised of five cyclic groups and five dihedral groups. A cyclic point group $C_n$ consists of all powers of an $n$-fold rotation while a dihedral point group $D_n$ contains $C_n$ as a subgroup and an additional two-fold in-plane rotation axis. In this paper we will focus on $C_n$ PGS.

With the point groups defined, we now consider the representations of point groups in a single particle Hilbert space. In the representation, every symmetry operation corresponds to an operator in the Hilbert space. As we will discuss both \emph{spinless} and \emph{spinful} fermions, it is necessary at this point to introduce double point groups, $G^D$. Mathematically speaking, $G^D=\{E,\bar{E}\}\otimes{G}$, where $E$ and $\bar{E}$ are the identify operator and a rotation of angle $2\pi$ about any axis in space, which satisfies $\bar{E}^2=E$, and $\bar{E}g=g\bar{E}$ for $g\in{G}$. This enlarged group has physical meanings as one considers its representations, which are classified into two types: in the first type $D(E)=D(\bar{E})=\mathcal{I}$, where $\mathcal{I}$ is the identity matrix; in the second type $D(E)=-D(\bar{E})=\mathcal{I}$. Representations of the first type are referred to as single-valued and representations of the second type are called double-valued. A degenerate subspace, in which every state has the same energy, of a single fermion must form a single-valued/double-valued representation of the underlying PGS of the Hamiltonian if the particles are spinless/spinful, as we know that a spinful fermion takes an extra minus sign as it is rotated by $2\pi$. For $C_n$ PGS, as one applies the general statement to Hilbert space representation of $C_n$ PGS, one can obtain $\hat{C}_n^n=(-1)^F\hat{I}$, and if in the orbital/band space representations, similarly, $\mathcal{C}_n^n=(-1)^F\mathcal{I}$, where $F=0$ ($F=1$) for spinless (spinful) fermions.

Given the Hilbert space representation $\hat{R}$ of a point group operator $R$, we now look at the sufficient and necessary conditions for a single particle tight-binding Hamiltonian $\hat{H}$ to be invariant under this operator. We assume that in one unit cell at $\mathbf{R}=n_1\mathbf{a}_1+n_2\mathbf{a}_2+n_3\mathbf{a}_3$ there are $s$ orbitals at $\mathbf{r}=\mathbf{R}+\mathbf{d}$, where $\mathbf{d}$ is the offset vector of the atom from the nearest lattice point. Note that we have ignored all unit cell structures, a simplification that spares us from discussing non-symmorphic space groups, which contain operations combined from member of a point group and translations by fractions of lattice vectors. Now the PGS operator $R$ sends an electron at $\mathbf{r}$ to $R\mathbf{r}=R\mathbf{R}+R\mathbf{d}$. If $R$ is a symmetry, there must be another atom at $R\mathbf{r}$, i.e., $\mathbf{R}'=R\mathbf{R}+R\mathbf{d}-\mathbf{d}$ is also a lattice point. Writing the above relation in second quantization, we have
\bea\label{eq:RealSpaceTransform}\hat{R}c_\alpha(\mathbf{R})\hat{R}^{-1}=\sum_{\beta}\mathcal{R}_{\alpha\beta}c_\beta(\mathbf{R}').\eea
In this equation, $\mathcal{R}_{\alpha\beta}$ describes a possible rotation of an orbital under $R$. For example, suppose we are considering $p_{x,y,z}$-orbitals under $R=C_4$ a four-fold rotation about $z$-axis, then obviously, $p_x$ is sent to $p_y$, $p_y$ to $-p_x$ and $p_z$ to $p_z$. In this case we have in the basis of $(p_x,p_y,p_z)$ \bea\mathcal{R}_4=\left(
                                                                                           \begin{array}{ccc}
                                                                                             0 & -1 & 0 \\
                                                                                             1 & 0 & 0 \\
                                                                                             0 & 0 & 1 \\
                                                                                           \end{array}
                                                                                         \right).\eea
If we adopt the following definitions of Fourier transform of $c_\alpha(\mathbf{R}+\mathbf{d})$
\bea c_\alpha(\bk)=\frac{1}{\sqrt{N}}\sum_\mathbf{R}c_\alpha(\mathbf{R})\exp(-i\bk\cdot\mathbf{R}),\eea
from Eq.(\ref{eq:RealSpaceTransform}), we obtain:\begin{widetext}
\bea\label{eq:kSpaceTransform}\hat{R}c_\alpha(\bk)\hat{R}^{-1}&=&\frac{1}{\sqrt{N}}\sum_\mathbf{R}\mathcal{R}_{\alpha\beta}c_\beta(\mathbf{R}')\exp(-i\bk\cdot\mathbf{R})\\
\nonumber&=&\frac{1}{\sqrt{N}}\sum_\mathbf{R}\mathcal{R}_{\alpha\beta}c_\beta(\mathbf{R}')\exp[-i\bk\cdot(R^{-1}\mathbf{R}'+R^{-1}\mathbf{d}-\mathbf{d})]\\
\nonumber&=&\frac{1}{\sqrt{N}}\sum_\mathbf{R}\mathcal{R}_{\alpha\beta}c_\beta(\mathbf{R}')\exp[-i(R\bk)\cdot(\mathbf{R}'+\mathbf{d}-R\mathbf{d})]\\
\nonumber&=&\mathcal{R}_{\alpha\beta}c_\beta(R\bk)e^{i\bk\cdot(\mathbf{d}-R\mathbf{d})}\\
\nonumber&\equiv&\bar{\mathcal{R}}_{\alpha\beta}(\bk)c_\beta(R\bk).\eea\end{widetext} It is easy to verify that $\bar{\mathcal{R}}(\bk)$ is a unitary matrix using that $\hat{R}$ is a unitary symmetry operation. A tight-binding model Hamiltonian with translational symmetry $\hat{H}=\sum_{\bk}\mathcal{H}_{\alpha\beta}(\bk)c^\dag_\alpha(\bk)c_\beta(\bk)$ transforms under $\hat{R}$ as
\bea\hat{R}\hat{H}\hat{R}^{-1}=\sum_{\bk}(\bar{\mathcal{R}}(\bk)\mathcal{H}(\bk)\bar{\mathcal{R}}^{-1}(\bk))_{\alpha\beta}c_\alpha(R\bk)c_\beta(R\bk).\eea
If $R$ is a symmetry, we have $\hat{R}\hat{H}\hat{R}^{-1}=\hat{H}$ therefore
\bea\bar{\mathcal{R}}(\bk)\mathcal{H}(\bk)\bar{\mathcal{R}}^{-1}(\bk)=\mathcal{H}(R\bk).\eea Using $\bar{\mathcal{R}}(\bk)=\mathcal{R}e^{i\bk\cdot(\mathbf{d}-R\mathbf{d})}$, we have
\bea\mathcal{R}\mathcal{H}(\bk)\mathcal{R}^{-1}=\mathcal{H}(R\bk)\label{eq:Rconstraint}.\eea Eq.(\ref{eq:Rconstraint}) determines the sufficient and necessary conditions for a single-particle tight-binding Hamiltonian to be invariant under a point group symmetry $R$. For a $C_n$ invariant 2D system, the above equation simplifies to
\bea\label{eq:Cnconstraint}&&\mathcal{C}_n\mathcal{H}(k_x,k_y)\mathcal{C}_n^{-1}=\\\nonumber&&\mathcal{H}(k_x\cos(\frac{2\pi}{n})-k_y\sin(\frac{2\pi}{n}),k_x\sin(\frac{2\pi}{n})+k_y\cos(\frac{2\pi}{n})),\eea where $\mathcal{C}_n$ is the transformation matrix in the orbital basis for the $n$-fold rotation about $z$-axis.

\subsection{Generalized definition of a topological insulator}

In this paper, we define a topological insulator as a system the many-body ground state of which cannot be adiabatically deformed into the ground state in the atomic limit\cite{ChenX2010,hughes2010inv}. The atomic limit is the limit of zero coupling between orbitals on different atoms but with all symmetries of the system preserved. Every system in this limit is considered as a trivial insulator. Clearly then, insulators that cannot be continuously tuned to the atomic limit while preserving the gap and all symmetries are nontrivial topological insulators according to this definition. This definition enables us to use any quantized quantity depending on the wavefunction to differentiate topologically trivial and non-trivial insulators. This is because a quantized quantity only changes when gap is closed, thus if the value of this quantity in an insulator is different than that in its atomic limit, it must be a topologically non-trivial insulator, or simply, a topological insulator. Such a quantity is called a topological index, the choice of which is determined completely by the dimensionality and the given symmetries of a Hamiltonian. This definition comprises more types of non-trivial topologies than those that may be labeled by a non-zero $Z$ or $Z_2$ index. Furthermore, it should be noted that a good quantum number that denotes a topological state is not necessarily a physical observable such as Hall conductance. Such a case has been noted previously in non-trivial topological insulators possessing inversion symmetry. These systems have topologically protected in-gap states at exactly $1/2$ in the entanglement spectrum and thus they cannot be removed by adiabatic deformations of the Hamiltonian\cite{hughes2010inv}. Nevertheless, while the number of in-gap states is a topological index, it is not related to any yet known experimental observable.

\subsection{Single Particle Entanglement Spectrum}
We have defined the general single particle entanglement spectrum in Eq.(\ref{eq:DefEntSpec}), and here we restrict the definition to the entanglement spectrum associated with a special type of subsystem: \emph{real space} cuts. To begin, the correlation matrix for a real space cut is given by
\bea\label{eq:Cdef}C_{i\alpha,j\beta}=\langle c^\dag_\alpha(\mathbf{r}(i))c_\beta(\mathbf{r}(j))\rangle,\eea where $\mathbf{r}(i)$ denotes a unit cell within a predefined subset of the system, called $A$, and $\alpha,\beta$ denote orbitals in a unit cell. Suppose in the selected subset one has $N_A$ unit cells and each unit cell has $N_{orb}$ orbitals, the entanglement is then given by a $N_A*N_{orb}\times N_A*N_{orb}$ matrix, where $N_A$ is the number of sites in the real space cut $A$. The set of eigenvalues of the matrix is defined as the single-particle entanglement spectrum or simply entanglement spectrum associated with subset $A$. The definition is a special one as it confines the $c$-operators within a predefined set of sites in real space. In general, there are other types of cuts such as momentum cuts or orbital cuts\cite{li2008}, which confine the $c$-operators within a range of momentum or a set of orbitals (e.g., Landau orbitals). From definition, it is easy to show that all eigenvalues of $C$, i.e., all `energies' in the spectrum, are within the range $[0,1]$. A cut may be chosen as the whole system, $L$, and one can prove that in this case the correlation matrix $C(L)$ must be a projector matrix, whose eigenvalue is either zero or unity\cite{hughes2010inv}.

The reason we are only interested in real space cuts in this paper is that we will focus on band insulators, whose constituent single particle wavefunctions are completely local in momentum space with $\bk$ as a good quantum number. Therefore all eigenvalues in a momentum cut must be either zero or unity, a situation usually referred to as trivial. On the other hand, a real space cut has a potential to distinguish insulators that can be adiabatically tuned into the atomic limit and those that cannot be. To better understand this point, let us consider an insulator in the atomic limit. In this limit all single particle wavefunctions are localized, and therefore when a real space cut is made, a particle must be either totally inside or totally outside the subsystem defined by the cut. This forces the eigenvalue of the correlation matrix to be either $1$ if the particle is located within the cut or $0$ if the particle is outside of the cut. If an insulator \emph{cannot} be adiabatically tuned to the atomic limit, \emph{then within the entanglement spectrum there must be eigenvalues that are topologically prevented from approaching arbitrary vicinities of $1$ or $0$}. We may gain additional insight into the entanglement spectrum by considering a 2D IQH state, or a Chern insulator, that has $C$ ($=$Chern number) protected extended states and, when a real space cut is made, there must be at least the same number of states that are neither inside nor outside the subsystem defined by the cut. These extended topological states present themselves as protected `in-gap', neither 0 nor 1, states in the entanglement spectrum. This means that existence of protected in-gap states in the entanglement spectrum implies existence of intrinsically delocalized states.

The single particle entanglement spectrum is closely related to the entanglement entropy\cite{kitaev2006,ryu2006,Ronny2010,Alexandradinata2011} by
\bea\label{eq:entropy} E_q=-\sum_n[\xi_n\ln\xi_n+(1-\xi_n)\ln(1-\xi_n)],\eea where $\xi_n$ is the $n$th eigenvalue in the spectrum. Entanglement entropy is known to quantify non-trivial topology in insulators: If the entanglement entropy cannot be adiabatically tuned to zero, the insulator is topologically non-trivial. When there are protected in-gap states in the entanglement spectrum, then from Eq.(\ref{eq:entropy}) the entanglement entropy always remain finite. Therefore, presence of protected in-gap states indicates non-trivial topology in an insulator.

\section{Entanglement Spectrum of Inversion Invariant Topological Insulators in 1D: A Revisit}
\label{sec:1Dinv}
\subsection{Decomposition of the correlation matrix into projectors}
Let us begin our study of the entanglement spectrum of point group symmetric topological insulators by examining the inversion invariant system\cite{hughes2010inv,Turner:2012} in 1D. The system is assumed to have $N$ sites where $N$ is an \emph{even} number. The inversion center is \emph{at the midway} of the $N/2$-th site and $(N/2+1)$-th site. For now, we assume that the system is translationally invariant, hence from Eq.(\ref{eq:Rconstraint}) we have $[\mathcal{P},\mathcal{H}(0)]=[\mathcal{P},\mathcal{H}(\pi)]=0$, by which we can label all eigenstates at $0$ and $\pi$ with their eigenvalues under $\mathcal{P}$ ($\pm1$). $k=0$ and $k=\pi$ are called the invariant $k$-points under inversion, denoted by $k_{inv}$. For a reason that is detailed in Appendix \ref{apndx:different_sewing}, we now use, instead of eigenvalues of $\mathcal{P}$, eigenvalues of $\bar{\mathcal{P}}(k)=\mathcal{P}e^{ik(d-Pd)}$ to label the eigenstates at $k=0,\pi$. For inversion invariant system, the symmetry fixed point is in the middle of two sites so $d=1/2$, and $d-Pd=1/2-(-1/2)=1$. From this we have $\bar{\mathcal{P}}(k)=e^{ik}\mathcal{P}$.

Now we separate the left-half of the system, from the $1$st to the $N/2$th site, as the subsystem $A$ and calculate the entanglement spectrum $\{\xi_n|A\}$. There can be eigenvalues in the spectrum lying exactly at $\xi_n=1/2$, and the number of these eigenvalues is given by \bea N_{1/2}=|\sum_{k_{inv}}(n^+(k_{inv})-n^-(k_{inv}))|\label{eq:1D_in-gap},\eea where $n^+(k_{inv})$ and $n^-(k_{inv})$ represent the number of \emph{occupied} states with $+1$ and $-1$ eigenvalue of $\bar{\mathcal{P}}(k_{inv})$ of the $m$th band, respectively. If there is an impurity potential, it will generally split the $1/2$-in-gap degeneracy; however, if the impurity potential is also inversion symmetric, having equal magnitude at the $i$-th site and the $(N+1-i)$-th site, the mid-gap states are preserved.

\begin{figure}[t]
\includegraphics[width=7cm]{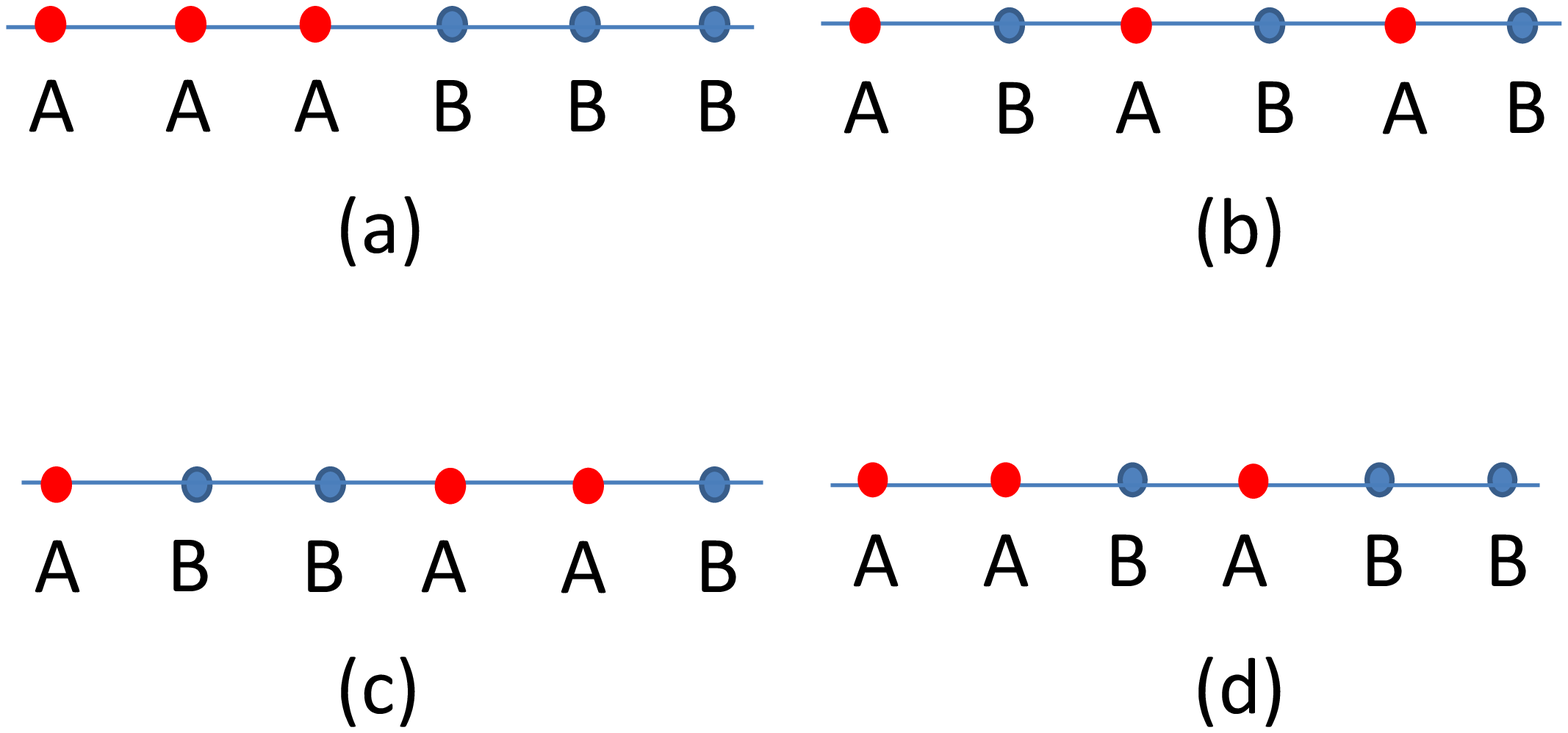}
\caption{Examples of inversion symmetric cut in a 1D system with six sites. All sites in red define the real space cut on which the entanglement spectrum is calculated.}\label{fig:oneDcut}
\end{figure}

In fact, we desire to make the statement more general by showing that the same result holds for all `inversion symmetric cuts', of which the above left-half cut is a special case. The subset which defines an inversion symmetric cut does not have to be the left or the right half, but should satisfy the following conditions: First, the subset $A$ has exactly $N/2$ sites, and second, if site $i$ is in $A$, site $N+1-i$ must not be in $A$. Such a subset $A$ will be referred to as an inversion symmetric cut. A set containing all sites left to the inversion center is a symmetric cut, and several other possible symmetric cuts are plotted in Fig.\ref{fig:oneDcut}. Below, we will use $r_i$ to denote the position of the $i$-th site in the cut for $i=1,...,N/2$. When the sum is over $i$, it means summing over $r_i$; when the sum is over $r$, it means summing over all sites in $L$.

While the correlation matrix $C$ in Eq.(\ref{eq:Cdef}) is not a projector, if $A$ is an inversion symmetric cut in an inversion invariant insulator, $C$ can be expressed as the linear average of two projectors $D$ and $\bar{D}$. To show this, we define new operators:\bea\label{eq:DefOfd}d_{i\alpha}&=&(c_\alpha(r_i)+\hat{P}c_\alpha(r_i)\hat{P})/\sqrt{2},\\
\nonumber \bar{d}_{i\alpha}&=&(c_\alpha(r_i)-\hat{P}c_\alpha(r_i)\hat{P})/\sqrt{2},\eea where $i=1,...,N/2$ and $\alpha$ is the orbital index (including spin). These are the symmetric and antisymmetric combinations of the annihilation operators. The single particle entanglement matrix for subsystem $A$ is:\bea C_{i\alpha,j\beta}&=&\langle{}c^\dag_{\alpha}(r_i)c_\beta(r_i)\rangle\\
\nonumber&=&\frac{1}{2}\langle{}(d^\dag_{i\alpha}+\bar{d}^\dag_{i\alpha})(d_{j\beta}+\bar{d}_{j\beta})\rangle.\eea Then we point out a simple fact:\bea\label{eq:orthogonality}\langle{}d^\dag_{i\alpha}\bar{d}_{j\beta}\rangle=\langle{}\bar{d}^\dag_{i\alpha}{d}_{j\beta}\rangle=0.\eea To prove this one only needs to notice that the ground state wavefunction must be an eigenstate of $\hat{P}$, and that $\hat{P}d_{i\alpha}\hat{P}=d_{i\alpha}$ while $\hat{P}\bar{d}_{i\alpha}\hat{P}=-\bar{d}_{i\alpha}$. Using Eq.(\ref{eq:orthogonality}), we have\bea C_{i\alpha,j\beta}=\frac{1}{2}(D_{i\alpha,j\beta}+\bar{D}_{i\alpha,j\beta}),\eea where\bea D_{i\alpha,j\beta}&=&\langle{}d^\dag_{i\alpha}d_{j\beta}\rangle,\\
\nonumber \bar{D}_{i\alpha,j\beta}&=&\langle{}\bar{d}^\dag_{i\alpha}\bar{d}_{j\beta}\rangle.\eea

Next we prove that $D$ and $\bar{D}$ are projector matrices, i.e., $D^2=D$ and $\bar{D}^2=\bar{D}$. In an inversion invariant system, all single particle eigenstates can be divided into two subspaces with even and odd parity respectively. We use $\psi_n$ to denote the annihilation operator of an even parity eigenstate and $\bar\psi_n$ to denote that of an odd parity eigenstate, i.e., $\hat{P}\psi_n\hat{P}=\psi_n$ and $\hat{P}\bar\psi_n\hat{P}=-\bar\psi_n$. Here $n$ is not a band label as before, because we do not have translational symmetry, but simply an index counting all single particle states. In a system with $N$ sites and $N_{orb}$ orbitals per site, there are $NN_{orb}/2$ states with even and odd parity respectively, so $n$ ranges from 1 to $NN_{orb}/2$ in general. Every creation/annihilation operator can be expressed in this basis and we have the unitary transform\bea c_{r\alpha}=U_{r\alpha,n}\psi_n+V_{r\alpha,n}\bar\psi_n.\eea Then from Eq.(\ref{eq:DefOfd}) we have $d_{r\alpha}=\sqrt{2}U_{r\alpha,n}\psi_n$. On the other hand we have \bea\psi_n&=&U^\dag_{n,r\alpha}c_{r\alpha},\\
\psi_n&=&\hat{P}\psi_n\hat{P}\\
\nonumber&=&U^\dag_{n,r\alpha}\hat{P}c_{r\alpha}\hat{P}\\
\nonumber&=&U^\dag_{n,r\beta}\mathcal{P}_{\beta\alpha}c_{N+1-r\alpha}.\eea Equating right hand sides of the last two equations, we have\bea\label{eq:eq21}U_{r\alpha,n}=\mathcal{P}_{\alpha\beta}U_{N+1-r\beta,n}.\eea

Using these relations we can prove that $D$ is a projector by calculating $D^2$ directly. In fact,\bea\label{eq:eq22}D_{j\beta,i\alpha}=\langle{}d^\dag_{j\beta}d_{i\alpha}\rangle=2\sum_{m\in{occ.}}U^\dag_{m,r_j\beta}U_{r_i\alpha,m}.\eea And we have\bea \label{eq:eq23}D_{j\beta,i\alpha}D_{l\gamma,j\beta}=4\sum_{m,n\in{occ.}}U_{r_i\alpha,m}U^\dag_{m,r_j\beta}U_{r_j\beta,n}U^\dag_{n,r_l\gamma},\nonumber\\\eea
where the summation is over the sites belonging to the $A$-subsystem.

Care should be taken to note that $U^\dag_{m,r_j\beta}U_{r_j\beta,n}\neq\delta_{mn}$! This is because by summing $j$ we only sum half of the system. When inserting Eq.(\ref{eq:eq21}) we obtain:\bea U^\dag_{m,r_j\beta}U_{r_j\beta,n}&=&U^\dag_{m,N+1-r_j\alpha}\mathcal{P}^\dag_{\alpha\beta}\mathcal{P}_{\beta\gamma}U_{N+1-r_j\gamma,n}\\
\nonumber&=&U^\dag_{m,N+1-r_j\beta}U_{N+1-r_j\beta,n}\;\textrm{[Do not sum over $j$].}\eea This leads to \bea\label{eq:eq25}U^\dag_{m,r_j\beta}U_{r_j\beta,n}=\frac{1}{2}U^\dag_{m,r\beta}U_{r\beta,n}=\frac{\delta_{mn}}{2}.\eea Substituting Eq.(\ref{eq:eq25}) into Eq.(\ref{eq:eq23}), we see $D^2=D$, or $D$ is a projector.

Using Eqs.(\ref{eq:eq22},\ref{eq:eq25}), it is easy to see that every row in $U_{r_j\alpha,n}$ is an eigenvector of $D$. If $\psi_n$ annihilates an occupied state, this vector has eigenvalue $1$ and if $\psi_n$ annihilates an unoccupied state, it has eigenvalue $0$. So the dimension of $D$, or the number of unity eigenvalues is: $dim(D)=Dim(\Psi_{occ.})$, or the number of occupied eigenstates with even parity. Similarly, it can be shown that $dim(\bar{D})=Dim(\bar{\Psi}_{occ.})$, or the number of occupied eigenstates with odd parity.

If $dim(D)>dim(\bar{D})$, there must be at least $dim(D)-dim(\bar{D})$ common eigenstates of $D$ and $\bar{D}$ with unity and zero eigenvalues, respectively. These states are therefore eigenstates of $C$ with eigenvalue exactly at $1/2$, i.e., $1/2$-in-gap states. And if $dim(D)<dim(\bar{D})$, there must be $dim(\bar{D})-dim(D)$ $1/2$-in-gap states. In general, we have $N_{1/2}=|dim(D)-dim(\bar{D})|.$ The two integers $dim(D)=Dim(\Psi_{occ})$ and $dim(\bar{D})=Dim(\bar{\Psi}_{occ})$ are the numbers of occupied eigenstates with even and an equal number of states with odd parities, or in other words, they are eigenstates in the two 1D representations of the inversion point group. These two numbers, $(z_1,z_2)=(Dim(\Psi_{occ}),Dim(\bar{\Psi}_{occ}))$, give a $Z^2$ classification of inversion invariant insulators. The sum of its two components $z_1+z_2$ is simply the total number of fermions, and their difference $|z_1-z_2|$ is exactly the number of protected in-gap states in the entanglement spectrum, i.e.,
\bea\label{eq:z1z2}N_{1/2}=|z_1-z_2|.\eea
In a trivial insulator, one has $z_1=z_2$, while any pair $(z_1,z_2)$ with $z_1\neq z_2$ denotes a topologically non-trivial inversion invariant insulator. Moreover, it should be noted that the proof presented here, up to this point, does not require translational symmetry.

\subsection{Translational invariance and number of in-gap states}

\begin{figure}
\includegraphics[width=8cm]{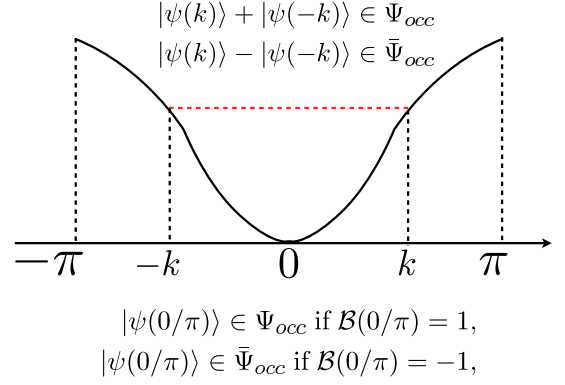}
\caption{A schematic of a single band in a 1D inversion symmetric insulator. From the figure we clearly see that at any $k=0$ we have two equal energy states at $k$ and $-k$, while at $k=0,\pi$, there is only one state. The two degenerate states at $\pm{k}$ can be recombined into two states with even and odd parities respectively, while the singlet state at $0$ or $\pi$ is either even or odd.}
\label{fig:1Ddispersion}
\end{figure}

Now we may proceed to calculate $Dim(\Psi_{occ})$ and $Dim(\bar{\Psi}_{occ})$ by relating them to the inversion eigenvalues at high symmetry $k$-points $0$ and $\pi$ in presence of translational symmetry. Suppose $\gamma_m(k)$ is the annihilation operator of the Bloch wavefunction at $k$ on the $m$th band, and define symmetry adapted operators
\bea\label{eq:eq26}\psi_m(k)&=&(\gamma_m(k)+\hat{P}\gamma_m(k)\hat{P})/\sqrt{2},\\
\nonumber\bar\psi_m(k)&=&(\gamma_m(k)-\hat{P}\gamma_m(k)\hat{P})/\sqrt{2},\eea and they satisfy $\hat{P}\psi_m(k)\hat{P}=\psi_m(k)$, $\hat{P}\bar\psi_m(k)\hat{P}=-\bar\psi_m(k)$. Due to inversion symmetry, the single particle states $|\psi(k)\rangle=\psi^\dag(k)|0\rangle$ and $|\bar\psi(k)\rangle=\bar\psi^\dag(k)|0\rangle$, in which $|0\rangle$ refers to the vacuum state and not the Fermi sea, are the eigenstates of $\hat{H}$. The two sets together, for $m\in{occ}$, make a basis of the filled bands in which every basic vector has a certain parity. The subspaces $\Psi^{(+1)}_{occ}$ and $\Psi^{(-1)}_{occ}$ are given by: $\Psi_{occ}=\{|\psi_m(k)\rangle|m\in{occ.},k\ge0,\psi_m(k)\neq0\}$ and $\bar\Psi_{occ}=\{|\bar\psi_m(k)\rangle|m\in{occ.},k\ge0,\bar\psi_m(k)\neq0\}$. We observe that we have a rather strange looking constraint $\psi_m(k)\neq0$ and $\bar\psi_m(k)\neq0$. This is because at $k=0,\pi$, either $\psi_m(k_{inv})=0$ or $\bar{\psi}_m(k_{inv})=0$, and either the state $|\psi_m(k)\rangle$ or $|\bar\psi_m(k)\rangle$ becomes a null state. To see this, we introduce the sewing matrix
\bea \mathcal{B}_{mn}(k)=\langle u_m(-k)|\hat{P}|u_n(k)\rangle\;{(m,n\in{occ})}.\eea At $k=0$ and $k=\pi$, Hamiltonian commutes with inversion, and, therefore, the sewing matrix must be diagonal with diagonal elements being either $+1$ or $-1$. Applying the sewing matrix by utilizing a property proven in Appendix \ref{apndx:sewing_property}, we have
\bea\label{eq:inversionsewing}\hat{P}\gamma_m(k)\hat{P}=\mathcal{B}_{nm}\gamma_n(-k).\eea Inserting the above equation into Eq.(\ref{eq:eq26}) and take $k=0,\pi$, we obtain
\bea\label{eq:1Dsewing}\psi_m(k_{inv})&=&(1+\mathcal{B}_{mm}(k_{inv}))\gamma_m(k_{inv})/\sqrt{2},\\
\nonumber\bar\psi_m(k_{inv})&=&(1-\mathcal{B}_{mm}(k_{inv}))\gamma_m(k_{inv})/\sqrt{2}.\eea It is then obvious that if $\mathcal{B}_{mm}(k_{inv})=1$, then $\bar\psi_m(k_{inv})=0$, otherwise $\psi_m(k_{inv})=0$. This is the mathematical description of the rather simple fact that for each at $k\neq0,\pi$, there is its counterpart at $-k$, from which we can construct two states with even and odd parities respectively, while at $k=0,\pi$ we cannot do this and either the even or the odd state must be a null state.
The counting of states in $\Psi_{occ}$ and $\bar{\Psi}_{occ}$ can be done on each band separately and then we add the numbers together. At $k\neq0,\pi$, $\psi_m(k)$ and $\bar\psi_m(k)$ appear in pairs so they contribute equally to $Dim(\Psi_{occ})$ and $Dim(\bar\Psi_{occ})$ (see Fig.\ref{fig:1Ddispersion} for an example). The number of states in $\Psi_{occ}$ and $\bar\Psi_{occ}$ contributed by this part is $(N-2)/2$ for each. Counting the states at inversion invariant points needs information of the sewing matrix.

To be more concrete, let us examine a system with only one band filled. Suppose we have $B(0)=B(\pi)=1$, using Eq.(\ref{eq:1Dsewing}), then\bea\psi(0)&=&=\sqrt{2}\gamma(0),\\
\nonumber\bar\psi(0)&=&0,\\
\nonumber\psi(\pi)&=&\sqrt{2}\gamma(\pi),\\
\nonumber\bar\psi(\pi)&=&0.\eea
Both at $k=0$ and $k=\pi$, there is one more state in $\Psi_{occ}$ than in $\bar\Psi_{occ}$. From this we see that $z_1=Dim(\Psi_{occ})=(N-2)/2+2$ and $z_2=Dim(\bar\Psi_{occ})=(N-2)/2$. According to previous discussion, we have $|z_1-z_2|=2$ $1/2$-in-gap states. One can repeat the discussion for $B(0)=B(\pi)=-1$ and find $z_1=Dim(\Psi_{occ})=(N-2)/2$ and $z_2=Dim(\bar\Psi_{occ})=(N-2)/2+2$. There are also two $1/2$-in-gap states for this case. If $B(0)=-B(\pi)$, we have $z_1=z_2=N/2$, which means no protected $1/2$-in-gap states.

For more than one filled band, we repeat the process for each filled band, and find that $z_1=Dim(\Psi_{occ})$ is the total number of $+1$ in the diagonal of $\mathcal{B}(0)$ and $\mathcal{B}(\pi)$, and $z_2=Dim(\bar\Psi_{occ})$ is the total number of $-1$ in the diagonal of $\mathcal{B}(0)$ and $\mathcal{B}(\pi)$:
\bea z_1&=&n_+(0)+n_+(\pi)=\sum_{k_{inv}}n_+(k_{inv})\\
\nonumber z_2&=&n_-(0)+n_-(\pi)=\sum_{k_{inv}}n_-(k_{inv}).\eea
Substituting these expressions into Eq.(\ref{eq:z1z2}), we recover Eq.(\ref{eq:1D_in-gap}). (The sewing matrix used here is different from the one used in Ref.[\onlinecite{hughes2010inv}] due to different choice of inversion center.)

\subsection{Protection of in-gap states against inversion symmetric and weak disorder}

We would like to understand how these new quantum numbers, $z_1$ and $z_2$ will change when we consider a disordered system. Since our previous discussion in Sec.\ref{sec:1Dinv}(A) did not rely on translational invariance the answer will not change so far as the disorders are inversion symmetric. We begin by assuming a form for the on-site potential which preserves the inversion symmetry $\hat{V}=\sum_{\alpha r}V(r)c^\dag_{\alpha}(r)c_{\alpha}(r)$ satisfying $V(r)=V(N+1-r)$, then we have $\hat{P}\hat{V}\hat{P}=\hat{V}$. Therefore,
\bea\langle\bar\psi_m|\hat{V}|\psi_n\rangle=0.\eea This entails that if $\hat{V}$ is tuned adiabatically, an even parity state $|\psi_m\rangle\in\Psi$ will always remain in the even parity subspace $\Psi$, so the total number of occupied states of even parity, $z_1$, does not change during the process. Similar argument gives that $z_2$ also remains unchanged during this adiabatic process. The many-body ground state for $\hat{H}+\hat{V}$ will be the same as the state evolved from the many-body ground state for $\hat{H}$, if there is no level crossing in the adiabatic process. Therefore, we conclude that, as far as $|V|$ is small compared with the bulk gap, the numbers of occupied states $z_1$ and $z_2$ in even and odd subspaces are the same as $z_1$ and $z_2$ in the homogeneous system. Thus, the number of $1/2$-in-gap states, $|z_1-z_2|$, is unchanged and the $1/2$-in-gap states are robust against any inversion symmetric disorder, and do not depend on translational symmetry in general.

Before ending our revisit to the 1D inversion invariant insulators, we must emphasize the following two points that give us insights for extending our work on inversion invariant insulators to $C_n$ invariant insulators. First, we want to choose a subsystem that is conjugal to its complementary subsystem with respect to the point group symmetry, because in this case, the correlation matrix decomposes into a linear average of projectors. Second, we can define an integer for each 1D representation of the point group, which is the total number of occupied states that transform according to that representation. The sum of these integers gives the total number of electrons, while their difference is related to the protected in-gap states in the entanglement spectrum.

\section{Entanglement Spectrum of Inversion Invariant Topological Insulators in 2D and 3D}
\label{sec:2D3Dinv}
The conclusion we arrived at for the 1D insulators with inversion symmetry can be easily generalized to higher dimensions. In these insulators, one can also define the $Z^2$ index as the number of occupied states with even and odd parities: $(z_1,z_2)=(Dim(\Psi_{occ}),Dim(\bar{\Psi}_{occ}))$. The sum of its two components $z_1+z_2$ gives the total number of electrons and their difference, $|z_1-z_2|$, is the number of $1/2$-in-gap states in the entanglement spectrum for any inversion symmetric cut. Similar to the symmetric cut in 1D, one defines a symmetric cut in 2D (3D) as a proper subset of sites called $A$ that has exactly one half of all sites, and if site $\mathbf{r}\in{A}$, then $P\mathbf{r}\bar{\in}A$. The detailed proof of $N_{1/2}=|z_1-z_2|$ is omitted because it takes exactly the same steps, which we simply summarize here as: 1. Write the correlation matrix $C$ as the linear average of two projector matrices $D$ and $\bar{D}$. 2. Show that the number of $1/2$-in-gap states is the difference between the number of unity eigenvalues of $D$ and $\bar{D}$. 3. Show that $dim(D)=Dim(\Psi_{occ})\equiv z_1$ and $dim(\bar{D})=Dim(\bar\Psi_{occ})\equiv z_2$.

We can easily count $Dim(\Psi_{occ})$ and $Dim(\bar\Psi_{occ})$ in presence of translational invariance. For each pair of $(\bk,-\bk)$ where $\bk\neq\bk_{inv}$, there are for each band exactly one parity odd and one parity even eigenstate, thus contributing equally to $Dim(\Psi_{occ})$ and $Dim(\bar\Psi_{occ})$. The contribution from all these pairs is $(N-4)N_{occ}/2$ in 2D and $(N-8)N_{occ}/2$ in 3D, as there are four and eight $\bk_{inv}$'s in 2D and 3D respectively. In the following, we will neglect the contribution from all of the aforementioned pairs in the $Z^2$ index, as they are completely determined by the total number of sites and the total number of bands (or particles). As in 1D, the interesting part is contributed by eigenstates at $\bk_{inv}$'s, because the eigenstate at $\bk_{inv}$ on each band is either parity even or parity odd. The contribution from these $\bk_{inv}$'s to the $Z^2$ index is given by\bea
Dim(\Psi_{occ.})&=&\sum_{\bk_{inv}}n^+(\bk_{inv}),\\
\nonumber Dim(\bar\Psi_{occ.})&=&\sum_{\bk_{inv}}n^-(\bk_{inv}).\eea The $Z^2$ index obtained in a translationally invariant system remains unchanged in the disordered system as long as the disorder potential preserve inversion symmetry, as mentioned previously. We may generally state that the number of $1/2$-in-gap states in the entanglement spectrum for any inversion symmetric cut with or without weak disorder is given by the same Eq.(\ref{eq:1D_in-gap}), with the only difference that now $\bk_{inv}$ runs over four and eight points in 2D and 3D respectively.

In addition to giving the number of $1/2$-in-gap states, the $Z^2$ index also gives the parity of Chern number in 2D. In fact, we have\bea(-1)^{(z_1-z_2)/2}&=&(-1)^{\sum_{\bk_{inv}}n^-(\bk_{inv})}\\
\nonumber&=&(-1)^C.\eea In the last equality, we have applied Eq.(92,93) in Ref.[\onlinecite{hughes2010inv}], which relate the Chern number to the inversion eigenvalues at all inversion invariant points in BZ.

In 2D and 3D, if in addition to inversion symmetry, we also have time-reversal symmetry, every band is at least doubly degenerate (for spinful fermions). And since $[\hat{T},\hat{P}]=0$, the two degenerate bands must have equal inversion eigenvalue at $\bk_{inv}$'s. In this case $(z_1-z_2)/2$ must be an even number and one can further have
\bea(-1)^{(z_1-z_2)/4}&=&(-1)^{\sum_{\bk_{inv}}n^-(\bk_{inv})/2}\\
\nonumber&=&(-1)^{\gamma_0},\eea where $\gamma_0$ is the $Z_2$ index of a TRI insulator.

As a final remark, the $Z^2$-index only concerns the total number of occupied states with even/odd parities and therefore does not exhaust the topological classifications. For example, an inversion symmetric insulator with only one filled band having $\mathcal{B}(\Gamma)=\mathcal{B}(X)=-\mathcal{B}(Y)=-\mathcal{B}(M)=1$ is topologically different from another one with $\mathcal{B}(\Gamma)=\mathcal{B}(M)=-\mathcal{B}(X)=-\mathcal{B}(Y)=1$, though they have the same $Z^2$-indices.

\section{Entanglement Spectrum in $C_n$ invariant Insulators}
\label{sec:Cn}

We are already aware that we can define a $Z^2$ index for inversion invariant insulators. This index represents the number of occupied eigenstates in each of the two 1D representations of the inversion point group. In total, there are $n$ 1D representations of point group $C_n$, in each of which $C_n$ is represented by a root of the equation $x^n=(-1)^F$. We denote the $m$th root by $x_m=e^{i(F+2(m-1))\pi/n}$. This allows us to define a $Z^n$ index $(z_1,...,z_n)$, in which $z_m$ is the number of occupied eigenstates in a subspace $\Psi^{(x_m)}$\bea z_m=Dim(\Psi_{occ}^{(x_m)}).\eea Any state $|\psi\rangle$ in $\Psi^{(x_m)}$ satisfies $\hat{C}_n|\psi\rangle=x_m|\psi\rangle$, or, in other words, transforms according to the $m$-th representation of group $C_n$.

Yet it is not clear in what manner this $Z^n$ index is related to the number of protected in-gap states in the single particle spectrum. To answer this, we must define what a symmetric cut is in a $C_n$ invariant insulator, analogous to the cut defined with inversion invariant insulators, because a given entanglement spectrum is associated with a specific real space cut. Due to the complexity of $C_n$ compared with the inversion point group, there are more than one type of symmetric cuts. Two integers $m_1$ and $m_2$, both being factors $n$, describe a symmetric cut denoted by $A^{m_2}_{1/m_1}$ that satisfies (a) the number of sites in $A$ is exactly $N/m_1$. (b) Subset $A$ is invariant under $m_2$-fold rotation, or symbolically, $C_{m_2}\mathbf{r}\in{A}$, $\forall\mathbf{r}\in{A}$ and (c) the subset $A$, together with all copies of $A$ obtained from acting $C_{m_1m_2}$ rotation on $A$ for $k<m_1$ times, constitute the whole system, or symbolically, $\cup_{k=0,...,m_1-1}C^k_{m_1m_2}A=L$, where $L$ represents the whole lattice. Notice that properties (a) and (c) together imply (d) $C^{k_1}_{m_1m_2}A\cap{C}^{k_2}_{m_1m_2}A=\emptyset$, if $k_1\neq k_2 \;\textrm{mod}\;m_1$, while $C_{m_1m_2}^{m_1}=C_{m_2}$. In other words, an $m_1m_2$-fold rotation on $A$ generates other equivalent subsystems that are invariant under $m_2$-fold rotation. For $n=2,3,4,6$, all possible combinations of $(m_1,m_2)$ are listed with typical examples of every type of cuts in the real space in Table \ref{tab:cuts}. Additionally, we note that for a fixed $n$, there is a one-to-one mapping between one type of symmetric cuts and one coset decomposition of a subgroup of $C_n$.

In the last column of Table \ref{tab:cuts}, the number of the protected in-gap states is given in terms of $Z^n$ indices. This gives the \emph{least} number of eigenvalues in the entanglement spectrum that are constrained in the range $[1/m_1,1-1/m_1]$. If $m_1=2$, they are all at exactly $1/2$ as we have seen in the inversion invariant insulators. But when $m_1>2$, we do not have protected states at exactly $1/2$, but states that are prevented from moving outside the range of $[1/m_1,1-m_1]$.

\begin{figure}
\includegraphics[width=8cm]{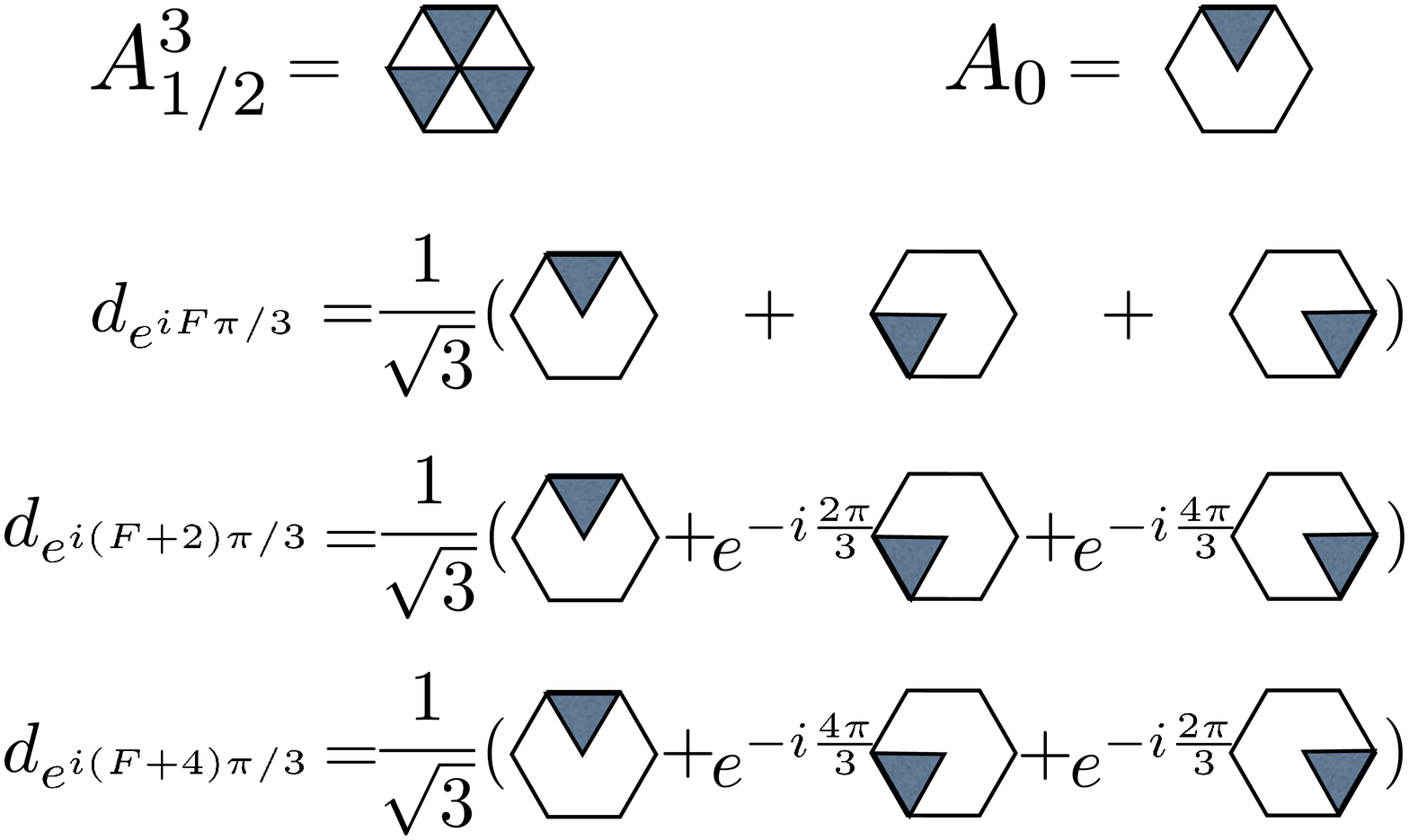}
\caption{A picturesque expression of the definition of operators $d_{y_q,\alpha}$. We use $A^3_{1/2}$ as an example, and show how the operators in $A$ can be decomposed into three sectors with different eigenvalues of $C_3$.}
\label{fig:eq38}
\end{figure}

To prove these expressions, we first show that the entanglement matrix calculated with an $A^{m_2}_{1/m_1}$ , $C(A^{m_2}_{1/m_1})$, can be block-diagonalized into $m_2$ blocks. Since $A$ is invariant under $C_{m_2}$, we can define a proper set of $A$ with exactly $N_A/m_2$ sites, $A_0$, such that $C^p_{m_2}\mathbf{r}{\notin}A_0$ if $\mathbf{r}\in{A_0}$ for $p=1,...,m_2-1$. For every site $\mathbf{r}\in{A_0}$ we define the following $m_2$ symmetry adapted operators\bea\label{eq:dyq}d_{y_q,\alpha}(\mathbf{r})=\frac{1}{\sqrt{m_2}}\sum_{p=1,...,m_2}(y_q^\ast)^{p}\hat{C}^p_{m_2}c_\alpha(\mathbf{r})\hat{C}^{-p}_{m_2},\eea where $y_q=\exp[i(2q+F-2)\pi/m_2]$ is the $q$th root of the $m_2$ roots of $e^{iF\pi}$ ($q=1,...,m_2$). (For an intuitive schematic of this definition, see Fig.\ref{fig:eq38}.) 

These new operators have two properties: $\{d_{y_q,\alpha}(\mathbf{r}),d^\dag_{y_{q'},\beta}(\mathbf{r}')\}=\delta_{qq'}\delta_{\alpha\beta}\delta_{\mathbf{r}\mathbf{r}'}$ and 
\bea\label{eq:drotation}
\hat{C}_{m_2}d_{y_q,\alpha}(\mathbf{r})\hat{C}_{m_2}^{-1}=y_q d_{y_q,\alpha}(\mathbf{r}).
\eea The first property ensures that the original operators $c_\alpha(\mathbf{r}\in{A})$ can be linked to $d_{y_q,\alpha}(\mathbf{r}\in{A}_0)$ by a unitary transform, thus preserving all eigenvalues of the correlation matrix. The second property leads to\bea\label{eq:orthogonal_d}\langle{d}^\dag_{y_{q_1},\alpha}(\mathbf{r})d_{y_{q_2},\beta}(\mathbf{r}')\rangle\propto\delta_{q_1q_2}.\eea To see this, we first notice that the ground state is $C_{m_2}$-invariant, i.e., $\hat{C}_{m_2}|0\rangle=e^{i\theta}|0\rangle$, where $\theta$ is some arbitrary angle. Then use Eq.(\ref{eq:drotation}):
\bea&&\langle{d}^\dag_{y_{q_1},\alpha}(\mathbf{r})d_{y_{q_2},\beta}(\mathbf{r}')\rangle\\
\nonumber&=&\langle0|\hat{C}^{-1}_{m_2}(\hat{C}_{m_2}{d}^\dag_{y_{q_1},\alpha}(\mathbf{r})\hat{C}^{-1}_{m_2})(\hat{C}_{m_2}d_{y_{q_2},\beta}(\mathbf{r}')\hat{C}^{-1}_{m_2})\hat{C}_{m_2}|0\rangle\\
\nonumber&=&y^{-1}_{q_1}y_{q_2}\langle{d}^\dag_{y_{q_1},\alpha}(\mathbf{r})d_{y_{q_2},\beta}(\mathbf{r}')\rangle.\eea If $q_1\neq{q_2}$, from the above equation, we have $\langle{d}^\dag_{y_{q_1},\alpha}(\mathbf{r})d_{y_{q_2},\beta}(\mathbf{r}')\rangle=0$.

From these we know that using the basis of $d_{y_q,\alpha}(\mathbf{r}\in{A}_0)$, the correlation matrix is block diagonalized into $m_2$ blocks, i.e.,\bea C(A)=C^{(1)}(A_0)\oplus{C}^{(2)}(A_0)...\oplus{C}^{(m_2)}(A_0),\eea where \bea C^{(p)}_{\mathbf{r}\alpha,\mathbf{r}'\beta}(A_0)=\langle d^\dag_{y_p,\alpha}(\mathbf{r})d_{y_p,\beta}(\mathbf{r}')\rangle.\eea Using this result, the study of in-gap states in $C(A)$ reduces to the study of in-gap states in each block $C^{(p)}(A_0)$. When we have calculated the number of in-gap states, $N_{mid}^{(q)}$, for each block, we can simply add up all these numbers and obtain\bea N_{mid}(A^{m_2}_{1/m_1})=\sum_{q=1,...,m_2}N_{mid}^{(p)}\eea for the total number of in-gap states.

From here we only focus on one block, $C^{(p)}(A_0)$, of the block diagonalized correlation matrix. In Appendix \ref{apndx:decomposition}, we prove that $C^{(p)}(A_0)$ can be written as a linear average of $m_1$ projectors
\bea\label{eq:eq46}C^{(p)}(A_0)=\frac{1}{m_1}(D^{(p)}_1+...+D^{(p)}_{m_1}),\eea where $D^{(p)}_r$ is a projector, projecting any state into a subspace $\Phi^{(p)}_r$, in which every state satisfies these two conditions: (i) it is an eigenvector of $\hat{C}_{m_2}$ of eigenvalue $y_p$, and (ii) it is an eigenvector of $\hat{C}_{m_1m_2}$ with eigenvalue $\lambda^{(p)}_r$. Notice that since $\hat{C}_{m_2}=\hat{C}_{m_1m_2}^{m_1}$, we have the constraint $(\lambda^{(p)}_r)^{m_1}=y_p$. Solve it and we have $\lambda^{(p)}_r=\exp[i\pi(F+2((r-1)m_2+p-1))/(m_1m_2)]$ for $r=1,...,m_1$. Given an eigenstate of $C_n$ of eigenvalue $x_{m=1,...,n}$, if it is in $\Phi^{(p)}_r$, using $\hat{C}_{m_1m_2}=\hat{C}_n^{\frac{n}{m_1m_2}}$ $x_m$ must satisfy $x_m^{\frac{n}{m_1m_2}}=\lambda^{(p)}_r$.

In Appendix \ref{apn:decomposition}, we prove that the number of non-zero eigenvalues of $D^{(p)}_r$ is\begin{widetext}
\bea\label{eq:DprInZn}dim(D^{(p)}_r)&=&Dim(\cup_{x_m}\{\Psi_{occ}^{(x_m)}|x_m^{n/(m_1m_2)}=e^{i2\pi((r-1)m_2+p-1+F/2)/(m_1m_2)}\})\\
\nonumber&=&\sum_{x_m^{n/(m_1m_2)}=e^{i2\pi((r-1)m_2+p-1+F/2)/(m_1m_2)}}z_m.\eea\end{widetext}
Recall that in inversion invariant insulators, the difference between $dim(D)$ and $dim(\bar{D})$ corresponds to the number of in-gap states; here we have a very similar relation. The block of the correlation matrix $C^{(p)}$ has
\bea N^{(p)}_{mid}(A_0)=\max_{i,j=1,...,m_1}|dim(D^{(p)}_i)-dim(D^{(p)}_j)|\eea eigenstates within the range $[1/m_1,1-1/m_1]$ (see Appendix \ref{apndx:range} for proof). Therefore the total number of in-gap states is the sum of the number of in-gap states in each block:\bea\label{eq:NmidCn} N_{mid}(A^{m_2}_{1/m_1})=\sum_{p=1,...,m_2}\max_{i,j=1,...,m_1}|dim(D^{(p)}_i)-dim(D^{(p)}_j)|.\nonumber\\\eea Substituting Eq.(\ref{eq:DprInZn}), which relates $dim(D^{(p)}_r)$ to $Z^n$-index, into Eq.(\ref{eq:NmidCn}), we obtain the expression of the total number of protected in-gap states for the symmetric cut $A^{m_2}_{1/m_1}$ as shown in the last column of Table.\ref{tab:cuts}.

\begin{table*}
\includegraphics[width=12cm]{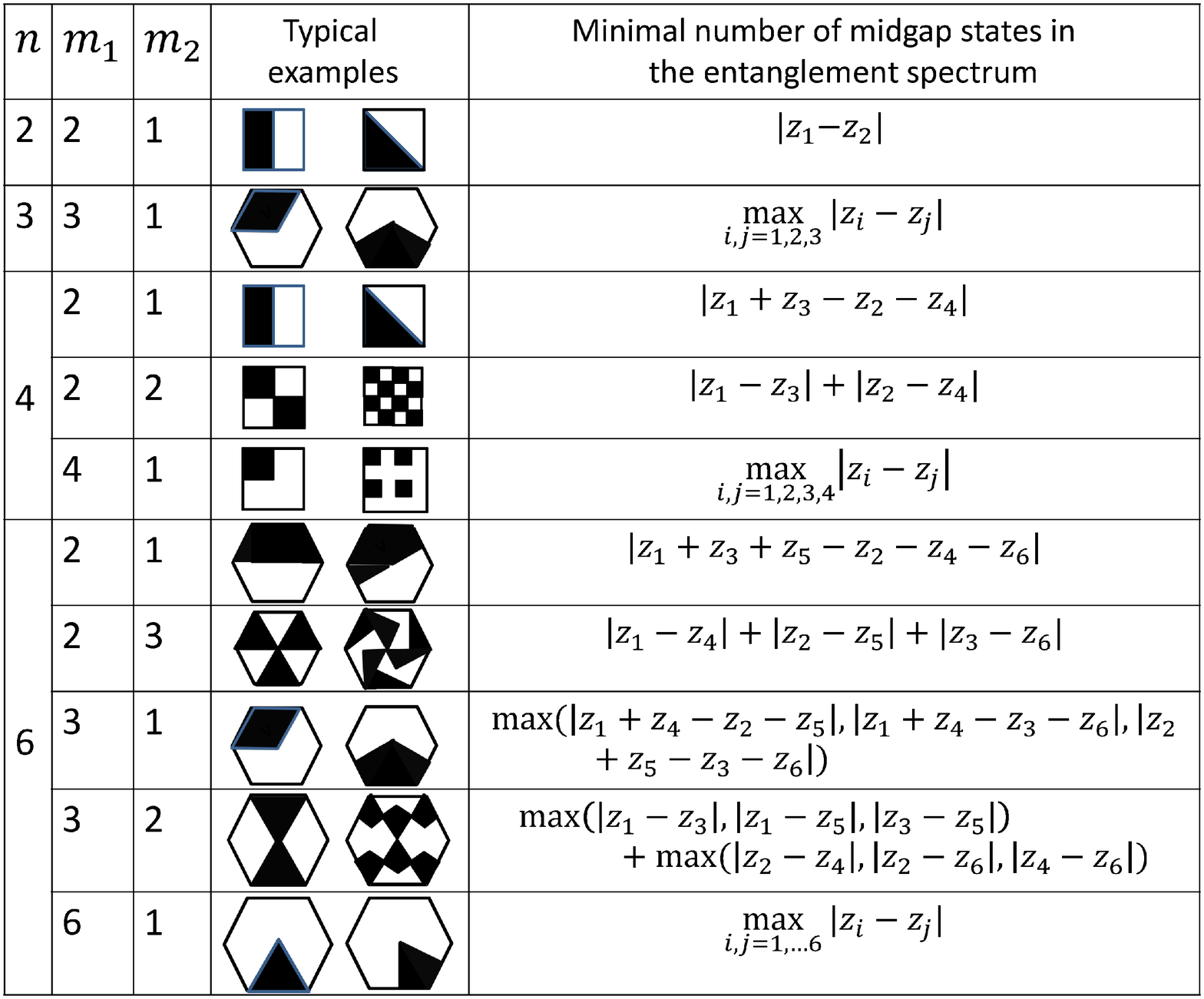}
\caption{All possible symmetric cuts $A_{1/m_1}^{m_2}$ in $C_n$ invariant insulators. In the fourth column, examples of each case are shown and in the last column, the minimal numbers of in-gap states in the entanglement spectrum are represented in terms of the $Z^n$-index.}
\label{tab:cuts}
\end{table*}

The discussion above applies to systems with and without translational invariance. In the case when translational invariance is present, we desire to elucidate how the $Z^n$ index related to the eigenvalues of symmetry operators at high symmetry points. To understand the nature of this link, we calculate the $Z^n$ index given a single particle Hamiltonian in $\bk$-space, or in the homogeneous limit with periodic boundaries. To be exact, we want to count the dimension of the linear subspace $\Psi_{occ}^{(x_m)}$, where $x_q=\exp(i(F+2q-2)\pi/n)$ for $q=1,...,n$. First we write symmetry adapted operators in terms of annihilation operators the Bloch states, $\gamma_m(\bk)$:
\bea\label{eq:DefOfPsiCn}\psi_{x_q,m}(\bk)=\frac{1}{\sqrt{n}}\sum_{p=0,...,n-1}(x_q^*)^p\hat{C}^{p}_n\gamma_m(\bk)\hat{C}^{-p}_n,\eea where $\gamma_m(\bk)$ is the annihilation operator of the Bloch state at $\bk$ on the $m$th band. This definition is a little different from that for $d_{y_q,\alpha}$ in Eq.(\ref{eq:dyq}), though they transform in similar ways under rotation: (i) $d_{y_q,\alpha}$ are to block diagonalize the correlation matrix, while $\psi_{x_q,m}$'s in general cannot do this; (ii) $d_{y_q,\alpha}(\br)$'s are \emph{not} annihilation operators of single particle eigenstates, while $\psi_{x_q,m}(\bk)$'s are generically annihilation operators of a single particle eigenstate; (iii) as we will see later, $\psi_{x_q,m}$ can be \emph{zero} at points of symmetry while $d_{y_q,\alpha}$ is always nonzero. We notice that for a generic $\bk$ that is not invariant under any subgroup of $C_n$, $\{\psi_{x_{q_1},m}(\bk),\psi^\dag_{x_{q_2},m'}(\bk)\}=\delta_{q_1q_2}\delta_{mm'}$. Any one of these these $\bk$ points on each band contributes $+1$ to every component of the $Z^n$ index. Again this contribution is ignored as it is completely determined by the total number of particles and total number of bands. But if $\bk_0$ is invariant under a subgroup $C_{n'}$ of $C_n$, i.e., $C_{n'}\bk_0=\bk_0$, we have, applying Eq.(\ref{eq:sewing3}) to Eq.(\ref{eq:DefOfPsiCn}),
\begin{widetext}
\bea\label{eq:psiandsewing}\psi_{x_q,m}(\bk_0)=\frac{1}{\sqrt{n}}\sum_{t=0,...,n'-1}({x_q^*}^{n/n'}(\mathcal{B}_{n'}(\bk_0))_{mm})^t\sum_{p=0,...,n/n'-1}(x_q^*)^p\hat{C}^{p}_n\gamma_m(\bk_0)\hat{C}^{-p}_n,
\eea
\end{widetext} where $\mathcal{B}_{n'}$ is the sewing matrix corresponding to the $n'$-fold rotation. This expression is non-zero if and only if $(\mathcal{B}_{n'}(\bk_0))_{mm}=x_q^{n/n'}$. It can be physically understood as follows: since $\hat{C}_{n'}=\hat{C}_n^{n/n'}$, and $(\mathcal{B}_{n'}(\bk_0))_{mm}$ is the eigenvalue of $\hat{C}_{n'}$ for the state at $\bk_0$, it must be the $n/n'$-th power of $x_q$ to be in the subspace of $\Psi^{(x_q)}$. Furthermore, in the BZ there are exactly $n/n'$ points that are $C_{n'}$ symmetric but not $C_n$ symmetric ($n'<n$), and therefore there are $n/n'$ orthogonal states, the linear combinations of which give the $n/n'$ eigenstates of $\hat{C}_n$ with eigenvalues satisfying $x_q^{n/n'}=(\mathcal{B}_{n'}(\bk_0))_{mm}$.

It is straightforward to show $\hat{C}_n\psi_{x_q,m}(\bk)\hat{C}_n^{-1}=x_q\psi_{\epsilon,m}(\bk)$. Therefore if $\psi_{x_q,m}(\bk)\neq0$ and $m\in{occ}$, then state $\psi^\dag_{x_q,m}(\bk)|0\rangle$ belongs to subspace $\Psi^{(x_q)}_{occ}$, contributing $+1$ to $z_q$. With Eq.(\ref{eq:psiandsewing}), the counting is therefore completely determined by the sewing matrices at these high symmetry points, or the number of each eigenvalue of $\hat{C}_{n'}$ at these points. The result for $n=2,3,4,6$ are listed in Table \ref{tab:dimensions}. In Appendix \ref{apn:explicit2}, we give an example of deriving the row with $n=6,m_1=2,m_2=3$ in Table \ref{tab:dimensions}, following the general rules given above.

\begin{table*}
\begin{tabular}{|c|c|c|}
\hline
$n=2$ & $Dim(\Psi_{occ}^{(e^{iF\pi/2})})$ & $n^{(e^{iF\pi/2})}_2(\Gamma)+n^{(e^{iF\pi/2})}_2(X)+n^{(e^{iF\pi/2})}_2(Y)+n^{(e^{iF\pi/2})}_2(M)$\\
\hline
    & $Dim(\Psi_{occ}^{(e^{i(F+2)\pi/2})})$ & $n^{(e^{i(F+2)\pi/2})}_2(\Gamma)+n^{(e^{i(F+2)\pi/2})}_2(X)+n^{(e^{i(F+2)\pi/2})}_2(Y)+n^{(e^{i(F+2)\pi/2})}_2(M)$\\
\hline
$n=3$ & $Dim(\Psi_{occ}^{(e^{iF\pi/3})})$ & $n^{(e^{iF\pi/3})}_3(\Gamma)+n^{(e^{iF\pi/3})}_3(K)+n^{(e^{iF\pi/3})}_3(K')$\\
\hline
    & $Dim(\Psi_{occ}^{(e^{i(F+2)\pi/3})})$ & $n^{(e^{iF\pi})}_3(\Gamma)+n^{(e^{i(F+2)\pi/3})}_3(K)+n^{(e^{i(F+2)/3\pi})}_3(K')$\\
\hline
    & $Dim(\Psi_{occ}^{(e^{i5F\pi/3})})$ & $n^{(e^{i(F+4)\pi/3})}_3(\Gamma)+n^{(e^{i(F+4)\pi/3})}_3(K)+n^{(e^{i(F+4)\pi/3})}_3(K')$\\
\hline
$n=4$ & $Dim(\Psi_{occ}^{(e^{iF\pi/4})})$ & $n_4^{(e^{iF\pi/4})}(\Gamma)+n_4^{(e^{iF\pi/4})}(M)+n_2^{(e^{iF\pi/2})}(X)$\\
\hline
    & $Dim(\Psi_{occ}^{(e^{i(F+2)\pi/4})})$ & $n_4^{(e^{i(F+2)\pi/4})}(\Gamma)+n_4^{(e^{i(F+2)\pi/4})}(M)+n_2^{(e^{i(F+2)\pi/2})}(X)$\\
\hline
    & $Dim(\Psi_{occ}^{(e^{i(F+4)\pi/4})})$ & $n_4^{(e^{i(F+4)\pi/4})}(\Gamma)+n_4^{(e^{i(F+4)\pi/4})}(M)+n_2^{(e^{iF\pi/2})}(X)$\\
\hline
    & $Dim(\Psi_{occ}^{(e^{i(F+6)\pi/4})})$ & $n_4^{(e^{i(F+6)\pi/4})}(\Gamma)+n_4^{(e^{i(F+6)\pi/4})}(M)+n_2^{(e^{i(F+2)\pi/2})}(X)$\\
\hline
$n=6$ & $Dim(\Psi_{occ}^{(e^{iF\pi/6})})$ & $n_6^{(e^{iF\pi/6})}(\Gamma)+n_3^{(e^{iF\pi/3})}(K)+n_2^{(e^{iF\pi/2})}(M)$\\
\hline
    & $Dim(\Psi_{occ}^{(e^{i(F+2)\pi/6})})$ & $n_6^{(e^{i(F+2)\pi/6})}(\Gamma)+n_3^{(e^{i(F+2)\pi/3})}(K)+n_2^{(e^{i(F+2)\pi/2})}(M)$\\
\hline
    & $Dim(\Psi_{occ}^{(e^{i(F+4)\pi/6})})$ & $n_6^{(e^{i(F+4)\pi/6})}(\Gamma)+n_3^{(e^{i(F+4)\pi/3})}(K)+n_2^{(e^{iF\pi/2})}(M)$\\
\hline
    & $Dim(\Psi_{occ}^{(e^{i(F+6)\pi/6})})$ & $n_6^{(e^{i(F+6)\pi/6})}(\Gamma)+n_3^{(e^{iF\pi/3})}(K)+n_2^{(e^{i(F+2)\pi/2})}(M)$\\
\hline
    & $Dim(\Psi_{occ}^{(e^{i(F+8)\pi/6})})$ & $n_6^{(e^{i(F+8)\pi/6})}(\Gamma)+n_3^{(e^{i(F+2)\pi/3})}(K)+n_2^{(e^{iF\pi/2})}(M)$\\
\hline
    & $Dim(\Psi_{occ}^{(e^{i(F+10)\pi/6})})$ & $n_6^{(e^{i(F+10)\pi/6})}(\Gamma)+n_3^{(e^{i(F+4)\pi/3})}(K)+n_2^{(e^{i(F+2)\pi/2})}(M)$\\
\hline
\end{tabular}
\caption{$Z^n$-index in a $C_n$ invariant insulator with translational symmetry. These indices are related to the number of all eigenvalues of the corresponding sewing matrices associated with subgroups of $C_n$ at high symmetry points within the BZ. $n_m^\epsilon$ represents the number of states with eigenvalue $\epsilon$ of $m$-fold rotation $C_m$, where $C_m$ is a subgroup of $C_n$.}
\label{tab:dimensions}
\end{table*}

Substituting the dimension counting into Eq.(\ref{eq:NmidCn}), we have finally linked the number of in-gap states in a $C_n$ invariant insulator with a symmetric cut $A_{1/m_1}^{m_2}$ to the number of different eigenvalues of $\hat{C}_m$ ($m$ dividing $n$) at all high symmetry points. Finally, it should be noted that although we use the sewing matrix, defined only for a homogeneous system, to calculate the $Z^n$ index, the components of the index do not change under any adiabatic shift of Hamiltonian including addition of $C_n$ symmetric disorder potential. Therefore, as far as the disorder strength is small compared with the insulating gap, the number of in-gap states given in Table \ref{tab:dimensions} is protected from changing only by $C_n$ symmetry but not by translational symmetry, and is uniquely determined by the $Z^n$ index.

\begin{figure*}
\includegraphics[width=18cm]{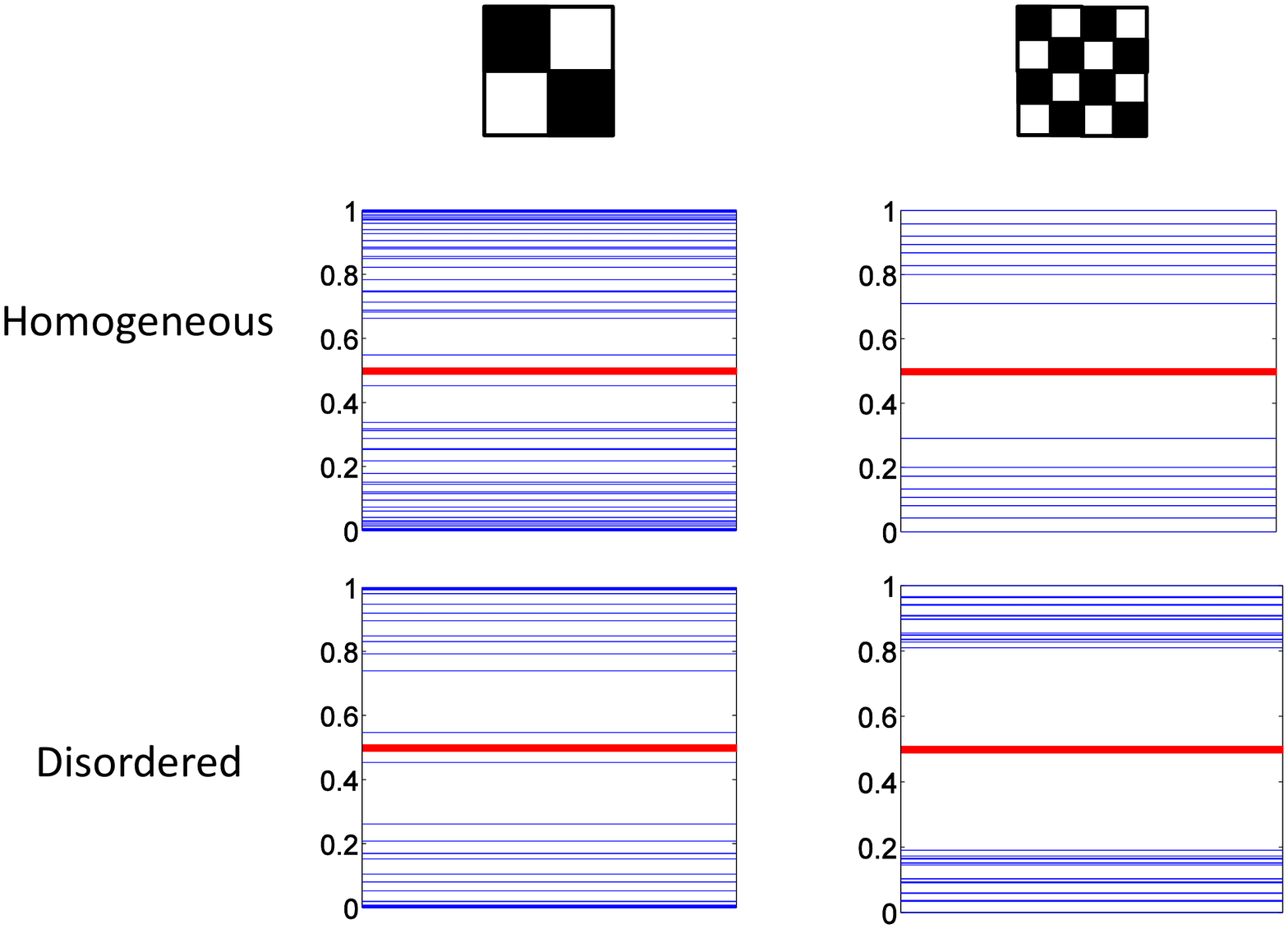}
\caption{Entanglement spectra in a $C_4$ invariant insulator described by the Hamiltonian in Eq.(\ref{eq:sample_C4_Ham}) with $m=1.5$ are plotted with two different cuts of the type $A^2_{1/2}$ on an 8-by-8 square lattice. The vertical axis is the entanglement eigenvalue and each level in the spectrum is represented by a horizontal line and  Different columns are for different specific cuts; the first row shows the result without impurity (with translational symmetry) and the second with random disorders of strength$\sim0.4$ that are $C_4$ symmetric. Red lines mark the position of eigenvalues at a value of exact $1/2$. The number of protected in-gap states is two in all cases.}
\label{fig:C4_ent_spec}
\end{figure*}

In Fig.\ref{fig:C4_ent_spec}, we give two specific examples of a $C_4$ invariant 2D insulator showing protected $1/2$-in-gap states in the entanglement spectrum, both with symmetric cuts of the type $A^2_{1/2}$. The system is described by a tight-binding model with two bands:\begin{widetext}\bea\label{eq:sample_C4_Ham}H(k_x,k_y)=(m-\cos(k_x)-\cos(k_y))\sigma_z+\sin(k_x)\sigma_x+\sin(k_y)\sigma_y,\eea\end{widetext} where $m$ is a real tunable parameter. This Hamiltonian is $C_4$-invariant, $C_4(\bk)H(k_x,k_y)C_4^{-1}(\bk)=H(-k_y,k_x)$, with $C_4(\bk)=\exp(i\sigma_z\pi/4)e^{ik_x}$. Mark that $C_4$-matrix is $\bk$-dependent since the rotation center is not chosen at a lattice site, but the center of a plaquette. Using $C_2(k_x,k_y)=C_4(-k_y,k_x)C_4(k_x,k_y)$, we have $C_2(\bk)=i\sigma_ze^{i(k_x-k_y)}$. We assume that the system is half-filled. There are two $C_4$-symmetric points $\Gamma$ and $M$, where the Hamiltonians are $H(\Gamma)=(m-2)\sigma_z$ and $H(M)=(m+2)\sigma_z$, respectively. If $m>2$, the occupied state is $(0,1)^T$ at both points, giving $\mathcal{B}(\Gamma)=e^{-i\pi/4}$ and $\mathcal{B}(M)=-e^{-i\pi/4}$. At $X$ or $Y$, $H(X/Y)=m\sigma_z$, and the occupied state is still $(0,1)^T$, giving $\mathcal{B}(X/Y)=i$. According to Table \ref{tab:dimensions}, we find $z_1=z_2=z_3=z_4=1$. For $m<-2$, similarly, we can similarly derive $z_{1,2,3,4}=1$. For $0<m<2$, at $\Gamma$, the occupied state becomes $(1,0)^T$, contributing $\mathcal{B}(\Gamma)=e^{i\pi/4}$, while for $M$ and $X$, the occupied state is still $(0,1)^T$. In this case we have $z_1=2$, $z_2=z_3=1$ and $z_4=0$. Finally for $-2<m<0$, one can similarly derive $z_1=z_4=1$, $z_2=2$ and $z_3=0$. Summarizing all possibilities, we have $z_1=1-sgn(m-2)/2+sgn(m)$, $z_2=1+sgn(m+2)/2-sgn(m)/2$, $z_3=1-sgn(m+2)+sgn(m)/2$ and $z_4=1+sgn(m-2)/2-sgn(m)$. From Table \ref{tab:cuts}, we know that with this type of symmetric cut, the number of $1/2$ in-gap states are $N_{1/2}=|z_1-z_3|+|z_2-z_4|=2|sgn(m+2)-sgn(m-2)|$. If $|m|>2$, then there is no protected in-gap state and if $|m|<2$, there are always two degenerate entanglement eigenvalues at exactly $1/2$. In Fig.\ref{fig:C4_ent_spec}, we confirm this result by calculating the entanglement spectrum with or without $C_4$ invariant impurities. In the top two figures of Fig.\ref{fig:C4_ent_spec}, we can see that although the details of the spectrum depends on the specific choice of the real space cut, the presence of the doubly degenerate protected in-gap states is common to both cuts. In the bottom two figures of Fig.\ref{fig:C4_ent_spec}, we can see that random impurities in general push the eigenvalues to the two ends of the spectrum as the electronic states become more localized, the doubly degenerate in-gap states always remain at $1/2$.

\section{Discussion}
\label{sec:discuss}
\subsection{Entanglement spectrum of 3D insulators with PGS}
The major focus of the paper is 2D insulators with $n$-fold rotation symmetry; however, the results obtained can be promptly extended to 3D insulators with $n$-fold rotation symmetry. This is because a 3D $C_n$ insulator can be considered as a 2D $C_n$ insulator parameterized by a continuous $k_z$ along the rotation axis. For every $k_z$ we can find its number of in-gap states in the entanglement spectrum and, as we require the 3D system to be insulating, this number of in-gap states must be a constant for different $k_z$. Therefore, studying the system at one $k_z$-slice as a 2D insulator is sufficient to understanding the symmetry protected in-gap states in the entanglement spectrum in the 3D $C_n$-invariant system.

For 3D insulators with other than $C_n$ PGS, their entanglement spectra in principle show other symmetry protected features, and the number of in-gap states depend on other factors. Nevertheless, if the underlying PGS has only 1D representations, as $C_n$ group does, one can still define the $Z^{g_0}$-index, where $g_0$ is the order of the group, similar to the $Z^n$-index for $C_n$-invariant insulators. Then one can make exact statements about the number of in-gap states for properly defined symmetric real space cuts. If the group has higher-dimensional irreducible representations, it is yet to be studied how the occupation numbers of states belonging to these representation are related to the entanglement spectrum.

\subsection{Effects of weak interaction}
Throughout the paper we have been studying systems without interactions. Now we discuss how weak interaction changes our results. By `weak', we mean that the ground state of the interacting system can be obtained by adiabatically transforming the Slater determinant state or the ground state of the non-interacting system. This implies that the interacting ground state is both gapped and non-degenerate. Therefore, we are not considering any symmetry breaking state as such a state must be degenerate with at least another state related by the broken symmetry. Additionally, we are not considering fractional quantum Hall systems the ground states of which are degenerate. In this work, we assume the interacting ground state to be an insulator with a non-degenerate ground state and a many-body gap.

In an interacting insulator, single particle states lose their meaning, but the many-body ground state is still a 1D representation of $C_n$, giving a $Z_n$ index. The relation between this $Z_n$ index and the $Z^n$ indices can be found by noticing that the $C_n$ rotation operator can be expressed as
\bea
\hat{C}_n=\exp[i\frac{2\pi}{n}\sum_{m=1,...,n,\alpha,\mathbf{r}\in{L/n}}(m-1)d^\dag_{x_m,\alpha}(\mathbf{r})d_{x_m,\alpha}(\mathbf{r})]\nonumber\\\eea where
\bea
d_{x_m,\alpha}(r)=\frac{1}{\sqrt{n}}\sum_{p=1,...,n}x_m^p\hat{C}_n^p{c}_\alpha(\br)\hat{C}_n^{-p},
\eea
and that \bea z_m=\langle\sum_{\alpha,\mathbf{r}\in{L/n}}d^\dag_{x_m,\alpha}(\mathbf{r})d_{x_m,\alpha}(\mathbf{r})\rangle.\eea Since the ground state is an eigenstate of $\hat{C}_n$, we have
\bea\langle\hat{C}_{n}\rangle&=&\exp(i\frac{2\pi}{n}z)\\
\nonumber&=&\exp(i\frac{2\pi}{n}\sum_m(m-1)z_m),\eea or \bea\label{eq:Zn_to_Zn}z=\sum_{m=1,...,n}(m-1)z_m\;\textrm{mod}\;n.\eea Eq.(\ref{eq:Zn_to_Zn}) shows how the $Z^n$ classification in a non-interacting system downgrades to the $Z_n$ classification in an interacting one. This indicates that a non-interacting insulator with a non-trivial $Z^n$ index may have a trivial $Z_n$ index in presence of weak interaction, and hence two topologically distinct non-interacting insulators may become indistinct as interaction is turned on, similar to the situation in superconductors\cite{TurnerA2011,FidkowskiL2011,Rosch2012}.

Beyond the basic discussion presented here, the effects of interactions in point group symmetric topological insulators have many unanswered questions. A particularly important question is how the effects of interactions present in the changes in the entanglement spectrum. It is not clear how the in-gap states in the entanglement spectrum evolve as one turns on the interaction. We leave this question and other questions as to the role of interactions to future work.

\section{Conclusion}\label{sec:conclusion}

In our analysis of the resultant entanglement spectrum in point group symmetric topological insulators, we have proven the relation between the $1/2$-in-gap states in the entanglement spectrum in an inversion invariant insulator and the inversion eigenvalues at inversion invariant points. This was accomplished by noticing that, given any inversion invariant non-interacting insulator, one can always find a complete set of single particle eigenstates, of both the single particle Hamiltonian and inversion operator, and the numbers of occupied eigenstates that are odd/even under inversion are good quantum numbers. This results in a $Z^2$-index of the insulator. The sum of these two integers gives the total number of particles and their difference is exactly the number of $1/2$-in-gap states in the entanglement spectrum. We have extended this result from inversion invariant insulators to 2D invariant insulators with $C_n$ invariance: The number of occupied states that transform according to each 1D representation of $C_n$ is a good quantum number, resulting in a $Z^{n}$-index. In a translationally invariant system, this $Z^n$-index can be calculated by counting the number of each $C_m$ ($m$ dividing $n$) eigenvalue at high symmetry points. In the entanglement spectrum, the number of protected in-gap states is shown completely determined by the $Z^n$ index with or without translational invariance in any $C_n$ invariant insulator. Specially, if and only if all components in the $Z^n$ index are equal, then the resultant number of protected in-gap states is zero. Finally, we briefly discussed how a similar analysis can be extended to 3D insulators with point groups having only 1D representations, and demonstrated that under weak interaction, the $Z^n$-index for a $C_n$ invariant insulator reduces to a $Z_n$-index.

\begin{acknowledgments}
CF thanks A. Alexandradinata for helpful discussions. MJG acknowledges support from the AFOSR under grant FA9550-10-1-0459 and the ONR under grant N0014-11-1-0728 and a gift the Intel Corporation. BAB was supported by NSF CAREER DMR- 095242, ONR - N00014-11-1-0635, Darpa - N66001-11- 1-4110 and David and Lucile Packard Foundation.  MJG and BAB thank the Chinese Academy of Sciences for generous hosting.
\end{acknowledgments}

\onecolumngrid
\begin{appendix}

\section{Subtleties in the definition of inversion operator in $\bk$-space\label{apndx:different_sewing}}

The relation between the number of in-gap states and the parities of occupied states at $k=0,\pi$ in a 1D inversion invariant insulator, Eq.(\ref{eq:1D_in-gap}), is completely consistent with Eq.(12) of Ref.\onlinecite{hughes2010inv}, although two expressions take different forms.

The difference lies in the definition of the inversion operator. While Ref.\onlinecite{hughes2010inv} uses a $k$-independent $\mathcal{P}$ as the matrix representation of inversion operator in $k$-space, in this paper we use a $k$-dependent $\bar{\mathcal{P}}(k)=\mathcal{P}e^{ik}$ as in Ref.[\onlinecite{Turner:2012}]. The factor $e^{ik}$ comes from the fact that the inversion center is mid-bond, instead of on-site. Therefore, at $k=\pi$, the even/odd parity states in our paper correspond to odd/even states in Ref.\onlinecite{hughes2010inv}.

In 2D, the inversion operator in $\bk$-space is $\bar{\mathcal{P}}(\bk)=\mathcal{P}e^{ik_x+ik_y}$ on a rectangular lattice, because the inversion center is at $(\mathbf{a}_1+\mathbf{a}_2)/2$; in 3D, we use $\bar{\mathcal{P}}(\bk)=\mathcal{P}e^{ik_x+ik_y+ik_z}$ as the inversion matrix in $\bk$-space, because the inversion center is set at $(\mathbf{a}_1+\mathbf{a}_2+\mathbf{a}_3)/2$. An advantage of using such definitions is that the formula for the number of in-gap states in all dimensions takes exactly the same form:
\bea N_{1/2}=|\sum_{\bk_{inv}}n_+(\bk_{inv})-n_-(\bk_{inv})|.\eea

\section{A basic property of the sewing matrix\label{apndx:sewing_property}}
In this Appendix we derive a simple property of a general sewing matrix associated with a PGS operation $R$ we have applied to prove Eq.(\ref{eq:inversionsewing}). The sewing matrix $\mathcal{B}_R(\bk)$ is defined as
\bea\mathcal{B}_{mn}(\bk)=\langle{u}_m(R\bk)|\hat{R}|u_n(\bk)\rangle,\eea where $m,n\in{occ}$. Multiply both sides by $|u_m(R\bk)\rangle$ then sum over $m\in{occ}$, we have
\bea\label{eq:sewing1}\mathcal{B}_{mn}|u_m(R\bk)\rangle=|u_m(R\bk)\rangle\langle u_m(R\bk)|\hat{R}|u_n(\bk)\rangle.\eea Since $\hat{R}$ is a symmetry of the Hamiltonian, if one $\hat{R}|u_n(\bk)\rangle$ must also be an occupied state at $R\bk$ if $n\in{occ}$, therefore $\sum_{m\in{occ}}|u_m(R\bk)\rangle\langle u_m(R\bk)|$ can be replaced by the identity in the occupied bands. Eq.(\ref{eq:sewing1}) becomes
\bea\label{eq:sewing2}\hat{R}|u_n(\bk)\rangle=\mathcal{B}_{mn}|u_m(R\bk)\rangle.\eea Then we substitute
\bea|u_n(\bk)\rangle&=&\gamma^\dag_n(\bk)|0\rangle,\\
\nonumber|u_m(R\bk)\rangle&=&\gamma^\dag_m(R\bk)|0\rangle\eea into Eq.(\ref{eq:sewing2}), and obtain
\bea\label{eq:sewing3}\hat{R}\gamma_n(\bk)\hat{R}^{-1}=\mathcal{B}_{mn}\gamma_m(R\bk).\eea

\section{Decomposition of a correlation matrix into projectors\label{apndx:decomposition}}\label{apn:decomposition}
In Sec. \ref{sec:Cn}, we have shown that the correlation matrix of a $A^{m_2}_{1/m_1}$ cut can be block diagonalized into $m_2$ blocks, each denoted by $C^{(s)}(A)$. In this appendix, we prove that this matrix can be further decomposed into the average of $m_1$ projectors: \bea C^{(s)}(A)=\frac{1}{m_1}(D^{(s)}_1+...+D^{(s)}_{m_1}).\eea We further prove that the number of unity eigenvalues in the projector $D^{(s)}_{i}$ equals the number of occupied states the $C_n$ eigenvalues $\lambda$'s of which satisfy $\lambda^{n/(m_1m_2)}=e^{i2\pi((i-1)m_2+s-1+F/2)/(m_1m_2)}$.

Let us start from recalling that $C^{(s)}_{i\alpha,j\beta}=\langle d^\dag_{y_s,\alpha}(\br_i)d_{y_s,\beta}(\br_j)\rangle$, where $\br_{i,j}\in{A_0}$. For each $d_{y_s,\beta}(\br_i)$, we further define $m_1$ new operators\bea f^{(s)}_{\epsilon,\alpha}(\br_i)=\frac{1}{\sqrt{m_1}}\sum_{p=0,...,m_1-1}(\epsilon^*)^pC_{m_1m_2}^pd^{(s)}_\alpha(\br_i){C_{m_1m_2}^\dag}^p,\eea where $\epsilon^{m_1}=\exp[i2\pi(\frac{s-1+F/2}{m_2})]$. There are in total $m_1$ $\epsilon$'s that satisfy this condition, which are denoted by $\epsilon_k=e^{i2\pi(km_2+s+F/2)/(m_1m_2)}$ for $k=1,...,m_1$, and are all eigenvalues of operator $\hat{C}_{m_1m_2}$. This is very similar to what we have done in the 1D inversion symmetric insulators (see Sec.\ref{sec:1Dinv}(A)): over there $d_\alpha(\br)$ and $\bar{d}_\alpha(\br)$ are combinations of $c_\alpha(\br)$ that are even and odd under inversion; here $f_{\epsilon,\alpha}^{(s)}(\br_i)$ are linear combinations of $d_{y_s,\alpha}(\br)$ with different eigenvalues of $C_{m_1m_2}$. These operators have two properties (a) $C_{m_1m_2}f^{(s)}_{\epsilon,\alpha}(\br_i)C^\dag_{m_1m_2}=\epsilon f^{(s)}_{\epsilon,\alpha}(\br_i)$ and (b) $\{f^{(s)}_{\epsilon,\alpha}(\br_i),{f^{(s)}}^\dag_{\epsilon',\beta}(\br_j)\}=\delta_{\epsilon\epsilon'}\delta_{\alpha\beta}\delta_{\br_i\br_j}$.
From the first property, one derives $\langle{f^{(s)}}^\dag_{\epsilon,\alpha}(\br_i)f^{(s)}_{\epsilon',\beta}(\br_j)\rangle=0$ if $\epsilon\neq\epsilon'$. The derivation is similar to that of Eq.(\ref{eq:orthogonal_d}), only that here we use the fact that the ground state is an eigenstate of $\hat{C}_{m_1m_2}$. Therefore the correlation matrix $C^{(s)}(A_0)$ decomposes in the form of Eq.(\ref{eq:eq46}) where\bea (D^{(s)}_k)_{i\alpha,j\beta}=\langle{f^{(s)}}^\dag_{\epsilon_k\alpha}(\br_i)f^{(s)}_{\epsilon_k\alpha}(\br_j)\rangle.\eea For simplicity, the superfix $(s)$ will be suppressed and always implied throughout the remainder of this Appendix.

Now we need to prove that for every $k=1,...,m_1$, matrix $D_k$ is a projector, i.e., $(D_k)^2=D_k$. First we notice that $C_{m_1m_2}$ is a subgroup of $C_n$ or $C_n$ itself, so each single particle eigenstate has an eigenvalue of $C_{m_1m_2}$. There are $m_1m_2$ eigenvalues of $\hat{C}_{m_1m_2}$, dividing the single particle Hilbert space into $m_1m_2$ sectors, each of which is denoted by $\Psi^{\epsilon}=\Psi^\epsilon_{occ}+\Psi^\epsilon_{unocc}$, where $\epsilon$ is an eigenvalue of $\hat{C}_{m_1m_2}$. Suppose that in the $\epsilon$-sector, all eigenstates are generated by $\psi^\dag_{\epsilon,q}$, with $q=1,...,N*N_{orb}/(m_1m_2)$. The operators $\psi^\dag_{\epsilon,q}$'s thus satisfy (a)$\hat{C}_{m_1m_2}\psi^\dag_{\epsilon,q}\hat{C}^\dag_{m_1m_2}=\epsilon\psi^\dag_{\epsilon,q}$ and (b) $\{\psi_{\epsilon,q},\psi^\dag_{\epsilon',q'}\}=\delta_{\epsilon\epsilon'}\delta_{qq'}$.

The previously defined operators $f_{\epsilon_k,\alpha}(\br_i)$ can be decomposed into a linear superposition of all single particle eigenstates:
\bea
f_{\epsilon_k,\alpha}(\br_i)=\sum_{\lambda}W^{(\lambda)}_{i\alpha,q}\psi_{\lambda,q},\;\br_i\in{A}_0.\eea But according to the property (a) of $f_{\epsilon_k,\alpha}(\br_i)$ and property (a) of $\psi_{\lambda,q}$, we have $W^{(\lambda)}\neq0$ only if $\lambda=\epsilon_k$, and therefore \bea
f_{\epsilon_k,\alpha}(\br_i)=W^{(\epsilon_k)}_{i\alpha,q}\psi_{\epsilon_k,q}.\label{eq:temp12}\eea One is reminded that $\br_i\in{A_0}$, which is $1/m_2$ of $A$ and is therefore $1/(m_1m_2)$ of the whole lattice, therefore $W^{(\epsilon)}$ is in general a $N*N_{orb}/(m_1m_2)$-dimensional unitary matrix. Before proceeding, one is reminded of the difference between the matrix $U$ introduced in Sec.\ref{sec:1Dinv}(A), which is a unitary transform between the \emph{unsymmetrized} $c_\alpha(\br)$'s and the \emph{symmetrized} single particle eigenstate operators, and the matrix $W$, which is a transform between the \emph{symmetrized} operators $f_{\epsilon,\alpha}$ and the \emph{symmetrized} single particle eigenstate operators. In general, $W$ is a block-diagonal matrix with each block labeled by its $\lambda$. Each block in a block-diagonalized matrix must be unitary, if the whole matrix is unitary.

Using Eq.(\ref{eq:temp12}), one can calculate $D_k$:
\bea\label{eq:temp15}
(D_k)_{i\alpha,j\beta}&=&\sum_{q,q'}{W^{(\epsilon_k)}}^\dag_{q,i\alpha}{W^{(\epsilon_k)}}_{j\beta,q'}\langle\psi^\dag_{\epsilon_k,q}\psi_{\epsilon_k,q'}\rangle\\
\nonumber&=&\sum_{q\in{occ}}{W^{(\epsilon_k)}}^\dag_{q,i\alpha}{W^{(\epsilon_k)}}_{j\beta,q}.\eea
And $D_k^2$:
\bea
\label{eq:temp13}(D^2_k)_{i\alpha,j\beta}&=&\sum_{m,\gamma}(D_k)_{i\alpha,m\gamma}(D_k)_{m\gamma,j\beta}\\
\nonumber&=&\sum_{q,q'\in{occ}}{W^{(\epsilon_k)}}^\dag_{q,i\alpha}{W^{(\epsilon_k)}}_{m\gamma,q}{W^{(\epsilon_k)}}^\dag_{q',m\gamma}{W^{(\epsilon_k)}}_{j\beta,q'}.\eea
Using the unitarity of $W^{(\epsilon)}$ we have\bea
{W^{(\epsilon_k)}}_{m\gamma,q}{W^{(\epsilon_k)}}^\dag_{q',m\gamma}=({W^{(\epsilon_k)}}^\dag W^{(\epsilon_k)})_{qq'}=\delta_{qq'}.\label{eq:temp14}\eea Substituting Eq.(\ref{eq:temp14}) into Eq.(\ref{eq:temp13}), we have
\bea
(D^2_k)_{i\alpha,j\beta}=\sum_{q\in{occ}}{W^{(\epsilon_k)}}^\dag_{q,i\alpha}{W^{(\epsilon_k)}}_{j\beta,q}=(D_k)_{i\alpha,j\beta}.\eea

From the explicit expression of Eq.(\ref{eq:temp15}), it is straightforward to see that for every $q\in{occ}$, the column vector ${W^{(\epsilon_k)}}^*_{i\alpha,q}$ is an eigenvector of $D_k$ of eigenvalue $1$ and for every $q\notin{occ}$, it is an eigenvector of eigenvalue zero. Therefore $dim(D^{(s)}_k)$ is equal to the number of occupied states with $C_{m_1m_2}$ eigenvalue $\epsilon_k$, or the number of occupied states with $C_n$ eigenvalue $\lambda$ that satisfies $\lambda^{\frac{n}{m_1m_2}}=\epsilon_k$.

\section{Number of in-gap eigenvalues of a matrix $C=\frac{1}{n}(D_1+...+D_n)$, where $D_i$ is a projector\label{apndx:range}}

In this Appendix, we prove that if a matrix $C$ can be written as the average of $n$ projectors, i.e., $C=1/n(D_1+...+D_n)$, then there must be at least $N_{mid}$ eigenstates, the eigenvalues of which, $\epsilon$, are in the range\bea\epsilon\in[1/n,1-1/n],\eea. The number of these states is given by\bea\label{eq:in-gap_number}N_{mid}=\max_{i,j=1,...,n}|Dim(D_i)-Dim(D_j)|,\eea where $Dim(D_i)$ is the number of nonzero eigenvalues of $D_i$. 

Before going on, the authors would like to point out that the proof is completely mathematical and has very little relation to the physics presented in the paper. It is provided here only for completeness of the work.

First we prove a lemma: Given two hermitian matrices $A$ and $B$, with eigenvalues $a_i$'s and $b_i$'s, consider their sum $M=A+B$, then for any eigenvalue of $A$, say $a_i$, there must be an eigenvalue of $M$ that satisfies $a_i+b_{min}\le m^*\le a_i+b_{max}$, where $b_{min}$ ($b_{max}$) is the minimum (maximum) eigenvalue of matrix $B$. We can always define $B'=B-(b_{min}+b_{max})/2I$, while $B$ and $B'$ have the same eigenstates, and here the maximum/minimum eigenvalue of $B'$ is $\pm(b_{max}-b_{min})/2\equiv\pm b_0$. So for any eigenstate of $M'=A+B'$ whose eigenvalue is between $a_i-b_0$ and $a_i+b_0$, the same state must also be an eigenstate of $M=A+B$ with eigenvalue between $a_i+b_{min}$ and $a_i+b_{max}$.

Denote the eigenstate of $A$ with eigenvalue $a$ with $|a\rangle$, then consider the quantity
\bea
\langle{a}|{B'}^2|a\rangle=\sum_n|y_j|^2{b'_j}^2\le{b}_0^2,
\eea where we have used the expansion of $B'$ in terms of its eigenstates $B'=\sum_jb'_j|b'_j\rangle\langle{b}_j|$ and the expansion of $|a\rangle$ in $|b'_j\rangle$, $|a\rangle=\sum_jy_j|b'_j\rangle$.

On the other hand, one can always decompose $|a_i\rangle$ in terms of eigenvectors of $M'$
\bea\label{eq:expansion}
|a_i\rangle=\sum_nx_n|m'_n\rangle,\eea where $|m'_n\rangle$ is an eigenvector of $M'$ with eigenvalue $m'_n$. We again calculate the quantity
\bea\label{eq:D5}
\langle{a}|{B'}^2|a\rangle&=&\langle{a}|(M'-A)^2|a\rangle\\
\nonumber&=&\langle{a}(M'-a)^2|a\rangle\\
\nonumber&=&\sum_n|x_n|^2(m'_n-a)^2\\
\nonumber&\le&b^2_0.
\eea 
Since $|x_n|^2\ge0$ for each $n$ and $\sum_n|x_n|^2=1$, there must be at least one $n^*$ that satisfies
\bea|m'_{n^*}-a|\le{b_0},\eea or
\bea a-b_0\le m_{n^*}\le a+b_0.\eea Therefore for we know there must be an eigenvalue of $M$, $m_{n^\ast}=m'_{n^\ast}+(b_{max}+b_{min})/2$ that is between $a_i+b_{min}$ and $a_i+b{max}$.

Consider the matrix $D=D_1+...+D_n$, define $A=D_1+D_2$ and $B=D_3+...+D_n$. If $dim(D_1)\neq dim(D_2)$, then there must be at least $\delta_{12}=|dim(D_1)-dim(D_2)|$ number of eigenstates of $A$ with eigenvalue $1$. On the other hand, since each $D_i$ is a projector, $b_{min}=0$ and $b_{max}=n-2$. Suppose $|\phi_{1,...,\delta_{12}}\rangle$ are the degenerate eigenstates of $A$ with eigenvalue $1$, and they make a subspace called $\Phi_A$. Take $|a_1\rangle=|\phi_1\rangle$, from the lemma, we can prove that there must be one eigenstate $|\lambda_1\rangle$ with eigenvalue between $[1,n-1]$. Then if $\delta_{12}>1$, one is guaranteed to have a state within $\Psi_A$ that is orthogonal to $|\lambda_1\rangle$. Take that state as $|a_2\rangle$. The decomposition of $|a_2\rangle$ does not contain $|\lambda_1\rangle$, or, $x_{\lambda_1}=0$. Therefore, for Eq.(\ref{eq:D5}) to hold, there must be another eigenstate of $M$, called $|\lambda_2\rangle$ that has eigenvalue between $[1,n-1]$. Then if $\delta_{12}>2$, one is guaranteed to have a state within $\Phi_A$ that is orthogonal to both $|\lambda_{1,2}\rangle$, called $|a_3\rangle$. Start from $|a_3\rangle$, one obtains another eigenstate of $M$ with eigenvalue between $[1,n-1]$. The process continues until one has exactly $\delta_{12}$ eigenstates of $M$ with eigenvalues $[1,n-1]$. Now, one is \emph{not} guaranteed to have a state in $\Phi_A$ that is orthogonal to all $|\lambda_{1,...,\delta_{12}}\rangle$. In this way, we have proved that there must be at least $|dim(D_1)-dim(D_2)|$ eigenstates of $D$ having eigenvalues between $1$ and $n-1$. It means there are $|dim(D_1)-dim(D_2)|$ number of protected in-gap state in the spectrum of $C=D/n$.

Notice that in the proof, $D_1$ and $D_2$ are arbitrarily chosen, and we can in principle choose any two $D_i$'s. This means the least number of protected in-gap states is\bea N_{mid}=\max_{i,j=1,...,n}|dim(D_i)-dim(D_j)|,\eea and these states are protected because their eigenvalues must stay within the range $[1/n,1-1/n]$ and can never approach 0 or 1 infinitely.

\section{Explicit derivation of the number of in-gap states for a $A^3_{1/2}$-cut in a $C_6$-invariant system}
\label{apn:explicit1}
In this Appendix we explicitly show how we derive the expression of the number of protected in-gap states in terms of the $Z^n$ index for a $A^3_{1/2}$ cut in a $C_6$-invariant insulator.

In this system, the index $z_{m=1,...,6}$ represents the number of occupied states in the $m$-th representation of $C_6$, that is, the states that have eigenvalue $\exp(i\pi(F+2m-2)/n)$ under $C_6$. According to the process sketched in Sec.\ref{sec:Cn}, first we can block diagonalized the correlation matrix $C(A^3_{1/2})$ into three blocks. Then the total number of in-gap states is simply the sum of the numbers of in-gap states in each block. In the first block describes the entanglement between states with $C_3$ eigenvalue $\exp(iF\pi/3)$ inside and outside $A$; the second block describes the entanglement between states with $C_3$ eigenvalue $\exp(i\pi(F+2)/3)$ inside and outside $A$; the third block describes the entanglement between states with $C_3$ eigenvalue $\exp(i\pi(F+4)/3)$ inside and outside $A$.

Next, we notice that states with the same $C_3$-eigenvalue, i.e., contributing to the same block, in general have two different $C_6$-eigenvalues. States have $C_3$-eigenvalue $\exp(iF\pi/3)$ may have $C_6$-eigenvalue $\exp(iF\pi/6)$ or $\exp(i(F+6)\pi/6)$; states with $C_3$-eigenvalue $\exp(i(F+2)\pi/3)$ may have $C_6$-eigenvalue $\exp(i(F+2)\pi/6)$ or $\exp(i(F+8)\pi/6)$; states with $C_3$-eigenvalue $\exp(i(F+4)\pi/3)$ may have $C_6$-eigenvalue $\exp(i(F+4)\pi/6)$ or $\exp(i(F+10)\pi/6)$. Therefore, an occupied subspace with a certain $C_3$-eigenvalue can be split into two subspaces with different $C_6$-eigenvalues. The number of in-gap states in each block is exactly equal to the difference between the number of states in these two $C_6$-subspaces.

According to this analysis, we have $N_{mid}^{(1)}=|z_1-z_4|$, $N_{mid}^{(2)}=|z_2-z_5|$ and $N_{mid}^{(3)}=|z_3-z_6|$. The total number of in-gap states is $N_{mid}=N_{mid}^{(1)}+N_{mid}^{(2)}+N_{mid}^{(3)}=|z_1-z_4|+|z_2-z_5|+|z_3-z_6|$.

\section{Explicit derivation of $Dim(\Psi_{occ}^{e^{i(F+4)\pi/6}})$ in Table \ref{tab:dimensions}}
\label{apn:explicit2}
In this Appendix, we use the example of counting the dimension of $\Psi_{occ}^{e^{i(F+4)\pi/6}}$ to illustrate the way we use to derive Table \ref{tab:dimensions}.

From Appendix \ref{apn:decomposition}, we know that $Dim(\Psi_{occ}^{e^{i(F+4)\pi/6}})$ is the number of occupied states whose $C_6$-eigenvalue is $e^{i(F+4)\pi/6}$. We have also shown that only high symmetry points contribute to the \emph{difference} of the dimensions of subspaces, and therefore we ignore the contribution from states at generic $\bk$'s.

At $\Gamma$, the highest symmetry is $C_6$, so the contribution to $Dim(\Psi_{occ}^{e^{i(F+4)\pi/6}})$ is simply the number of states with $C_6$ eigenvalue $e^{i(F+4)\pi/6}$. At $K$, the highest symmetry is $C_3$. Notice that $C_3=C_6^2$, and we can see that if the $C_3$-eigenvalue of a state at $K$ is $e^{i(F+4)\pi/3}$, then it will contribute $+1$ to $Dim(\Psi_{occ}^{e^{i(F+4)\pi/6}})$. Finally, $M$ is a $C_2$-invariant point. Notice that $C_2=C_6^3$, and we can conclude that a state at $M$ contributes $+1$ to $Dim(\Psi_{occ}^{e^{i(F+4)\pi/6}})$ if and only if its $C_2$-eigenvalue is $\exp(iF\pi/2)$.

\end{appendix}
\twocolumngrid
%

\end{document}